# From Reversible Quantum Microdynamics to Irreversible Quantum Transport [*]

J. Rau [a] and B. Müller [b]

[a] *Max-Planck-Institut für Kernphysik, Postfach 103980, 69029 Heidelberg, Germany*

[b] *Physics Department, Duke University, Durham, NC 27708–0305, U.S.A.*

**Abstract**

The transition from reversible microdynamics to irreversible transport can be studied very efficiently with the help of the so-called projection method. We give a concise introduction to that method, illustrate its power by using it to analyze the well-known rate and quantum Boltzmann equations, and present, as a new application, the derivation of a source term which accounts for the spontaneous creation of electron-positron pairs in strong fields. Thereby we emphasize the fundamental importance of time scales: only if the various time scales exhibited by the dynamics are widely disparate, can the evolution of the slower degrees of freedom be described by a conventional Markovian transport equation; otherwise, one must account for finite memory effects. We show how the projection method can be employed to determine these time scales, and how –if necessary– it allows one to include memory effects in a straightforward manner. Finally, there is an appendix in which we discuss the concepts of entropy and macroscopic irreversibility.

# 1 Introduction

**Quantum transport equations confront us with both technical and conceptual problems.** The macroscopic evolution of complicated quantum systems can often be described by means of transport equations. Prominent examples are the Langevin equation, Master equation, hydrodynamic equations, rate equations, or the quantum Boltzmann equation. In this context the following technical questions arise:

– Under which conditions are such transport equations applicable? Can one formulate precise mathematical criteria for their validity?
– Can quantum transport equations be derived –in a rigorous fashion– from the underlying microscopic dynamics? And if so, how?

Furthermore, the macroscopic evolution usually displays a high degree of irreversibility, which is reflected in a continuous increase of entropy. Associated with these features are the following conceptual problems:

– What is the proper definition of the entropy? Why does its increase correspond to a physical thermalization of the system?
– How can the irreversible behavior be reconciled with the reversibility of the underlying microscopic dynamics?

**In principle, these problems have long been solved.** For the derivation of quantum transport equations there are a variety of tools available, such as the projection method [1–17], the –closely related– Zubarev method [18–20], thermofield theory [21–23], Wigner function [24,25], and Green's function techniques [26–29]. Independent of the particular tool employed, one generally arrives at the conventional transport equations only after certain approximations: coarse-graining, smoothing of rapidly oscillating functions, neglecting memory effects, to name just a few. All these approximations essentially amount to exploiting a separation of *scales* in the system [30,31]. Consequently, mathematical criteria for the validity of transport equations can usually be formulated in terms of these different scales. Such exploitation of widely disparate scales is not unique to quantum transport theory; it can already be found in classical transport theory. A famous example is furnished by the derivation of the classical Boltzmann equation: Boltzmann's "Stoßzahlenansatz" (assumption that collisions may be considered statistically independent) crucially depends on the fact that the duration of each individual scattering process is much shorter than the average time that elapses between two successive collisions. There are hence two time scales, $\tau_{\text{coll}}$ and $\tau_{\text{free}}$, which must be widely disparate: $\tau_{\text{coll}} \ll \tau_{\text{free}}$. Not surprisingly, the simple Boltzmann description breaks down as soon as this separation condition is no longer satisfied, as is the case in very dense or very strongly interacting plasmas.

The separation of scales, in conjunction with a large number of degrees of freedom, is also responsible for the irreversible features of the macroscopic dynamics. The conceptual issues surrounding this macroscopic irreversibility are by now well understood; all relevant ideas have at some point been spelt out and can be found in various places in the literature [32–36].

**Nevertheless, some of the issues have recently received renewed attention.** It is impossible to present here an exhaustive overview over all the current activities in the field of quantum transport; yet our own area of specialization alone –the theoretical study of highly energetic nucleus-nucleus collisions– has recently witnessed many investigations of questions like the following:

– Is the Boltzmann equation (or any other Boltzmann-like transport equation) applicable to high energy nucleus-nucleus collisions [37]?
– More specifically, are there non-negligible memory effects [38–40]? If so, how do they affect the evolution of the system? Can they be accounted for in a non-Markovian generalization of the Boltzmann equation? How can such a non-Markovian Boltzmann equation be derived from first principles?
– How can genuine quantum phenomena without classical analogues, such as spontaneous particle creation in a strong external field, be incorporated into existing kinetic equations [41–45]? How do they contribute to thermalization [46,47]?

Aside from these technical questions, controversies continue over the origin and meaning of irreversibility [48]. Some of the recent debate has been stimulated by work on "decoherence" [49–56], i.e., the phenomenon that macroscopic systems obey essentially classical equations of motion despite the quantum mechanical nature of the underlying microscopic dynamics.

**One tool that appears especially suited for the study of such issues is the projection method.** In order to be suitable for the investigation of the above issues, a theoretical framework should

– permit an analysis of time scales;
– if the scales are widely disparate: provide a well-defined, controlled procedure to exploit this separation and to recover conventional Markovian transport theories;
– if the scales are not widely disparate: allow for the derivation of non-Markovian transport equations and hence for the inclusion of finite memory effects;

- furnish a well-defined procedure to systematically expand any such non-Markovian correction in powers of time scale ratios; and
- permit the definition of a relevant entropy and thus the quantitative study of irreversible behavior.

One framework that satisfies all these conditions is the so-called projection method. As its name indicates, it is based on projecting the evolution of a quantum system onto some lower-dimensional subspace of the space of observables (Liouville space). The projection method is broadly applicable and affords a systematic way of deriving quantum transport equations from first principles. With its help, one may obtain many seemingly unrelated equations such as the Langevin-, Master-, Boltzmann-, or time-dependent Hartree-Fock equation. The approach permits a systematic analysis of time scales and thus furnishes the desired criteria for the validity of the Markovian limit. Furthermore, the projection method allows for the incorporation of finite memory effects and thus constitutes a useful tool for the study of non-Markovian corrections. Since it is formulated in the language of density matrices, mathematical manipulations and approximations within this formalism have a direct probabilistic, and therefore physical, interpretation, providing useful guidance during complicated calculations. Crucial concepts like the level of description, memory time, relevant entropy, etc., can be easily defined, making the projection method especially suited for the issues surrounding the irreversibility of macroscopic processes.

**Our review will focus on specific aspects of the projection method.** The projection method was originally developed by Nakajima [1], Zwanzig [2–6], Mori [7,8] and Robertson [9–13]. Its basic concepts have been reviewed [14] and are also presented in some textbooks on statistical mechanics [15–17]. Despite this wealth of available material we think that there are important reasons why another review might be called for: While in some branches of physics such as solid state physics [57] or quantum optics [58] the projection method has been employed extensively, there are other areas such as the physics of high energy nuclear collisions [59] in which quantum transport theory is used to describe the nonequilibrium dynamics of many-body systems, yet in which the projection method has found little or no application. In parts of the community, therefore, the projection method and its potential advantages have remained largely unknown.

For the reader who is not familiar with the projection technique we intend to give a concise introduction which focuses on the basic concepts, yet contains all essential formulae needed for applications. In order to demonstrate the power of the method we will present derivations –from first principles– of two well-known quantum transport equations, the rate equation for elastic scattering and the quantum Boltzmann equation. We will further illustrate the use of the projection method in linear response theory by deriving Drude's formula for the electrical conductivity. Aside from this general introduction and illustration we shall place particular emphasis on the following three aspects:

(i) *Time scales.* We will discuss
   - why time scales play a fundamental role in the description of macroscopic dynamics;
   - how the projection method can be used to determine these time scales (time scale analysis);
   - how, in this fashion, one can obtain quantitative criteria for the validity of conventional Markovian transport theories;
   - how –in cases in which these criteria are not fulfilled– the projection method allows one to derive generalized transport equations which are no longer Markovian and hence account for finite memory effects; and finally,
   - how such non-Markovian corrections may be expanded systematically in powers of time scale ratios.
   
   These aspects will be illustrated in the cases of the rate and quantum Boltzmann equations.
(ii) *New application.* We will derive a (generally non-Markovian) source term in the quantum Boltzmann equation which accounts for the spontaneous creation of particle-antiparticle pairs in strong fields.
(iii) *Entropy and irreversibility.* Even though –as we have claimed– all relevant ideas are already on the market, they are rarely presented in a coherent fashion. We wish to offer our own view of the subject, and discuss how the generally problematic issues (definition of relevant entropy, origin of irreversibility, etc.) present themselves in the context of the projection approach. In particular, we will emphasize the fundamental importance of disparate time scales and elaborate on the close connection between irreversibility and reproducibility of macroscopic experiments.

Our goal is to elucidate these specific areas of interest, rather than to attempt a complete account of all actual or potential applications of the projection method (a task which would appear quite Herculean). For this reason we offer our apologies in advance to those numerous practitioners of the projection method who might not find their work adequately represented.

**The review is divided into five parts:**

(i) Following this introduction we first collect some theoretical preliminaries, notably the concepts of Liouville space, superoperators and generalized canonical states (section 2).
(ii) Section 3 is then devoted to the essential ideas and general formalism of the projection method. We introduce the concept of projection su-peroperators and derive the general projected equation of motion for an arbitrary set of selected observables. We discuss two particularly important realizations of this projection which give rise to, respectively, the Langevin-Mori and Robertson equations. We also include a discussion of possible approximations, with special emphasis on the Markovian limit.

(iii) Next, in section 4, we turn our attention to specific examples and applications. After some general remarks on quantum transport equations we present detailed investigations of the acceleration term in the quantum Boltzmann equation, the rate equation for elastic scattering, the Drude formula for the electrical conductivity, the source term in the quantum Boltzmann equation which accounts for spontaneous particle creation, and the collision term in the quantum Boltzmann equation. In all cases we proceed as follows: we first employ the projection method to derive the exact *non*-Markovian equation of motion; then perform a thorough time scale analysis; obtain criteria for the validity of the Markovian limit; and only then, where appropriate, recover the conventional Markovian transport equation.

(iv) In section 5 we briefly summarize our findings and present an outlook on further potential applications of the projection method.

(v) There is an appendix devoted to conceptual issues. In that appendix we begin with a general investigation of the entropy concept, and discuss why the entropy represents an appropriate measure to describe a system's approach to equilibrium. We shall argue that entropy –along with the so-called "relative entropy"– serves mainly as a diagnostic tool, a claim that we support with an account of various uses of the entropy concept in statistical physics. Only then we proceed to discuss temporal variations of the entropy, in particular the second law of thermodynamics and the $H$-theorem. We emphasize the intimate connection between irreversibility, reproducibility, and the existence of well-separated time scales.

## 2 Preliminaries

### 2.1 Observables and Their Dynamics

**Observables are vectors in Liouville space.** Linear operators acting on Hilbert space can be added and multiplied by complex scalars; they may thus be regarded as elements of an abstract vector space. This vector space of linear operators is called *Liouville space* [17]. Its Hermitian elements, i.e., Hermitian operators $A, B, \ldots$, constitute *observables*. The *state* at time $t$ of a quantum system is described by the statistical operator, or density matrix, $\rho(t)$ with properties

$$\rho^\dagger = \rho \quad , \quad \rho \geq 0 \quad , \quad \operatorname{tr} \rho = 1 \quad . \tag{1}$$

In the state $\rho(t)$ an observable $A$ has the *expectation value*

$$\langle A \rangle(t) = \operatorname{tr}[\rho(t) A] \quad . \tag{2}$$

Liouville space is endowed with two scalar products: the *Liouville space inner product*,

$$(A|B) := \operatorname{tr}[A^\dagger B] \quad , \tag{3}$$

and the *canonical* (or *symmetrized*) *correlation function* [60] with respect to the state $\rho$,

$$\langle A; B \rangle_\rho := \int_0^1 d\lambda \operatorname{tr}\left[\rho^\lambda A^\dagger \rho^{1-\lambda} B\right] \quad . \tag{4}$$

The canonical correlation function with respect to a canonical equilibrium state $\rho = Z^{-1} \exp(-\beta H)$ is also known as the *Mori product*. In terms of these scalar products, expectation values can be written as

$$\langle A \rangle = (\rho|A) = \langle 1; A \rangle_\rho \quad . \tag{5}$$

**The dynamics is determined by the Liouville-von Neumann equation.** The dynamics of a quantum system is governed by its Hamiltonian $H$. In the Schrödinger picture the statistical operator evolves according to the *Liouville-von Neumann equation*

$$\frac{d\rho(t)}{dt} = -\frac{i}{\hbar}[H, \rho(t)] \quad . \tag{6}$$

This equation has the formal solution

$$\rho(t) = U(t, t_0)\rho(t_0)U^\dagger(t, t_0) \tag{7}$$

with $U(t, t_0)$ being the unitary *evolution operator*. A state $\rho$ is called *stationary* if $[H, \rho] = 0$; while an observable $A$ is called a *constant of the motion* if $[H, A] = 0$. Unless explicitly stated otherwise, we will always assume that neither the Hamiltonian nor the observables are explicitly time-dependent.

**The Liouville-von Neumann equation can be formulated and solved in the language of superoperators.** An operator $\mathcal{O}$ acting linearly on vectors in Liouville space, $A \to \mathcal{O}A$, is called a *superoperator*. Three important examples are the Liouvillian $\mathcal{L}$, the evolution superoperator $\mathcal{U}$ and the Green's function superoperator $\mathcal{G}$. The *Liouvillian* is defined as

$$\mathcal{L}A := \hbar^{-1}[H, A] \quad . \tag{8}$$

With its help the Liouville-von Neumann equation can be written in the form

$$\frac{d\rho}{dt} = -i\mathcal{L}\rho \quad , \tag{9}$$

while the equation of motion for expectation values takes the form

$$\frac{d}{dt}\langle A \rangle = i(\rho|\mathcal{L}A) \quad . \tag{10}$$

The Liouvillian is Hermitian with respect to the Liouville space inner product $(\cdot|\cdot)$, but generally not with respect to the canonical correlation function $\langle \cdot ; \cdot \rangle_\rho$ (unless $\rho$ is stationary).

The *evolution superoperator* is defined as

$$\mathcal{U}(t, t_0)\rho(t_0) := U(t, t_0)\rho(t_0)U^\dagger(t, t_0) \quad . \tag{11}$$

It determines –at least formally– the evolution of expectation values via

$$\langle A \rangle(t) = (\rho(t_0)|\mathcal{U}(t_0, t)A) \quad . \tag{12}$$

The evolution superoperator is unitary with respect to the Liouville space inner product, but in general not with respect to the canonical correlation function. It has the property

$$\mathcal{U}^{-1}(t, t_0) = \mathcal{U}(t_0, t) \quad , \tag{13}$$

and it satisfies the equation of motion

$$\frac{\partial}{\partial t}\mathcal{U}(t,t_0) = -i\mathcal{L}\mathcal{U}(t,t_0) \tag{14}$$

or

$$\frac{\partial}{\partial t}\mathcal{U}(t_0,t) = i\mathcal{U}(t_0,t)\mathcal{L} \quad, \tag{15}$$

respectively, with initial condition $\mathcal{U}(t_0,t_0) = 1$. Multiplication with a step function yields the *causal* evolution superoperator

$$\mathcal{U}_<(t_0,t) := \mathcal{U}(t_0,t) \cdot \theta(t-t_0) \tag{16}$$

(where '<' means '$t_0 < t$').

Finally, the *(causal) Green's function superoperator* is defined as the Fourier transform of the (causal) evolution superoperator,

$$\mathcal{G}_{(<)}(t_0,\omega) := \int_{-\infty}^{\infty} \mathrm{d}t \, \exp(i\omega t)\mathcal{U}_{(<)}(t_0,t) \quad. \tag{17}$$

Provided the Liouvillian $\mathcal{L}$ is time-independent and hence

$$\mathcal{U}(t_0,t) = \exp[i\mathcal{L}\cdot(t-t_0)] \quad, \tag{18}$$

the associated Green's operators are given by

$$\mathcal{G}_{(<)}(t_0,\omega) = \exp(i\omega t_0)\,\mathcal{G}_{(<)}(\omega) \tag{19}$$

with

$$\begin{aligned}\mathcal{G}(\omega) &:= 2\pi\delta(\omega+\mathcal{L}) \quad, \\ \mathcal{G}_<(\omega) &:= \left.\frac{i}{\omega+i\eta+\mathcal{L}}\right|_{\eta\to 0^+} \quad.\end{aligned} \tag{20}$$

Causal and ordinary Green's function superoperators are related via

$$\mathcal{G}_<(\omega) = \tfrac{1}{2}\mathcal{G}(\omega) + i\,\mathrm{Pr}\left(\frac{1}{\omega+\mathcal{L}}\right) \quad, \tag{21}$$

where 'Pr' denotes the principal value. With respect to the Liouville space inner product or any equilibrium correlation function, $\mathcal{G}(\omega)$ is Hermitian, while $[i \operatorname{Pr} \ldots]$ is anti-Hermitian.

## 2.2 Generalized Canonical States

**Generalized canonical states contain information about selected observables only.** At any given time $t$ let the expectation values of some selected observables $\{G_a\}$,

$$g_a(t) := \langle G_a \rangle(t) \quad , \tag{22}$$

be the only information available. This information is generally not sufficient to reconstruct unambiguously the statistical operator $\rho(t)$; there are many possible choices. However, one choice of particular importance is the *generalized canonical statistical operator*

$$\rho_{\text{can}}[g_a(t)] := Z(t)^{-1} \exp(-\lambda^a(t) G_a) \quad . \tag{23}$$

Here $Z(t)$ denotes the *partition function*

$$Z(t) := \operatorname{tr} \exp(-\lambda^a(t) G_a) \quad , \tag{24}$$

a summation over $a$ is implied (Einstein convention), and the time-dependent *Lagrange parameters* $\{\lambda^a(t)\}$ are adjusted so as to yield, at each time $t$, the correct expectation values for the selected observables.

Among all states which satisfy the constraints $\operatorname{tr}\rho = 1$ and $\operatorname{tr}(\rho G_a) = g_a$, the generalized canonical state is the one which maximizes the von Neumann entropy

$$S[\rho] := -k \operatorname{tr}(\rho \ln \rho) \quad . \tag{25}$$

For this reason the generalized canonical state may be considered "least biased" or "maximally non-committal" with regard to missing information. A more detailed discussion of this maximum entropy rationale can be found in the appendix.

**Expectation values, Lagrange parameters and entropy satisfy thermodynamic identities.** The following relations hold *at each time $t$*. The expectation values $\{g_a\}$ of the selected observables can be obtained from the partition function via

$$g_a = -\frac{\partial}{\partial \lambda^a} \ln Z \quad . \tag{26}$$

Inserting the generalized canonical state into the definition of the entropy yields

$$S = k \ln Z + k\lambda^a g_a \quad , \tag{27}$$

with differential

$$dS = k\lambda^a dg_a \quad . \tag{28}$$

Arbitrary expectation values $\langle A \rangle$ in the generalized canonical state $\rho_{\mathrm{can}}$ depend on the Lagrange parameters $\{\lambda^a\}$. If these vary infinitesimally, the expectation value $\langle A \rangle$ changes according to

$$d\langle A \rangle = -\langle \delta G_a; A \rangle d\lambda^a \quad . \tag{29}$$

Here the correlation function $\langle \cdot ; \cdot \rangle$ is evaluated in the state $\rho_{\mathrm{can}}$, and the observable $\delta G_a$ is defined as

$$\delta G_a := G_a - \langle G_a \rangle \quad . \tag{30}$$

With the help of the *correlation matrix*

$$C_{ab} := \langle \delta G_a; \delta G_b \rangle = -\frac{\partial g_b}{\partial \lambda^a} \tag{31}$$

which relates infinitesimal variations of $\lambda$ and $g$,

$$dg_b = -d\lambda^a C_{ab} \quad , \quad d\lambda^a = -dg_b (C^{-1})^{ba} \quad , \tag{32}$$

one may expand the expectation value also in $dg$:

$$d\langle A \rangle = (C^{-1})^{ba} \langle \delta G_a; A \rangle dg_b \quad . \tag{33}$$

# 3 Projection Method

## 3.1 *Motivation*

**The dynamics is projected onto the level of description.** When dealing with a complicated quantum system such as a gas or plasma, one is typically neither interested in, nor capable of, describing the time evolution of all its microscopic properties. Rather, two typical problems are

  (i) in linear response theory: to determine dynamical susceptibilities, which, according to Kubo's formula [61,62], amounts to calculating dynamical correlations of the form $\langle G_a; \mathcal{G}_<(\omega) G_b \rangle_{\text{eq}}$; and
 (ii) more generally, away from equilibrium: to determine the evolution of certain selected expectation values $\{\langle G_a \rangle(t)\}$.

Both calculations require knowledge of the dynamics of only a small set of selected ("relevant") observables $\{G_a\}$. These relevant observables, together with the unit operator, span a (relatively small) subspace of Liouville space, the so-called *level of description*.

The full statistical operator generally contains a lot of information which is not at all related to the relevant observables. This irrelevant information complicates calculations without providing any insight into the dynamics of the relevant observables. To discard such unnecessary "baggage," one *projects* the motion of the system onto the level of description. This projection yields a closed equation of motion for the relevant observables only; which, however, is generally no longer Markovian.

**The projection method allows one to exploit a separation of time scales.** Typically, neither the original Liouville-von Neumann equation nor the projected, non-Markovian equation of motion for the relevant observables can be solved exactly. One must resort to clever approximations, a task which, in practice, amounts to finding suitable expansion parameters. Two possibilities come to mind:

  (i) The Liouvillian may have the form $\mathcal{L} = \mathcal{L}^0 + \alpha \mathcal{V}$ with $\mathcal{V}$ denoting a perturbation and $\alpha$ a small coupling constant. In this case one may expand in powers of $\alpha$ (perturbation expansion).
 (ii) Macroscopic systems may exhibit widely separated time scales $\tau_{\text{short}}$ and $\tau_{\text{long}}$. In this case an additional expansion parameter is furnished by the ratio $(\tau_{\text{short}}/\tau_{\text{long}})$. As will be shown later, the zero-th order of such an expansion corresponds to the *Markovian limit*.

Table 1
Examples for sets of relevant observables, associated relevant parts of the statistical operator, and kinds of discarded information

| Physical system | Relevant observables $\{G_a\}$ | Relevant part $\rho_{\rm rel}(t)$ | Discarded information |
|---|---|---|---|
| $\Sigma^1 \times \Sigma^2 \times \ldots$ several subsystems | all $\Sigma^i$-observables ($i = 1, 2, \ldots$) | $\rho^1(t) \otimes \rho^2(t) \otimes \ldots$ | correlations between subsystems |
| many-particle system | all or selected single-particle observables | $Z^{-1} \exp(-\sum \lambda^{ij}(t) a_i^\dagger a_j)$ (Hartree form) | particle correlations |
| $S \times B$ system in heat bath | all $S$-observables and bath energy | $\rho^s(t) \otimes Z^{-1} \exp(-\beta(t) H^b)$ | internal degrees of freedom of the heat bath; system-bath correlations |
| arbitrary | all observables commuting with $A = \sum_i a_i P_i$ | $\sum_i P_i \rho(t) P_i$ | coherence with respect to $A$ |

While the original Liouville-von Neumann equation may lend itself to a perturbative treatment, it is not a good starting point for an expansion in powers of time scale ratios. In contrast, the projected equation of motion –with the influence of irrelevant degrees of freedom mapped onto a non-local behavior in time– allows one to detect and systematically exploit a separation of time scales. This is the basic practical merit of the projection method.

*3.2 General Formalism*

*3.2.1 Relevant Part of the Statistical Operator*

**Discarding irrelevant information yields the relevant part of the statistical operator.** The relevant observables have time-dependent expectation values $g_a(t) \equiv \langle G_a \rangle(t)$. Given these expectation values as constraints, one may at all times $t$ define an associated canonical state $\rho_{\rm can}[g_a(t)]$. Due to the entropy maximization involved, it contains information pertaining to the relevant observables only; irrelevant information has been discarded entirely. For this reason the generalized canonical state is also called the *relevant part of the statistical operator* and denoted by $\rho_{\rm rel}(t)$. Some common examples for sets of relevant observables, associated relevant parts of the statistical operator, and kinds of discarded information, are listed in table 1.

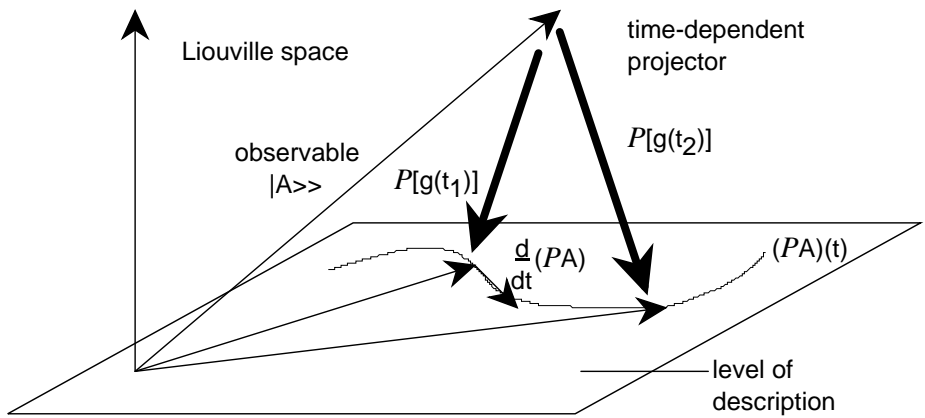

Fig. 1. Arbitrary observables are projected onto the level of description. The projection may vary in time.

*3.2.2 Projectors*

**Arbitrary observables are projected onto the level of description.**
A projector in Liouville space is a superoperator $\mathcal{P}$ satisfying $\mathcal{P}^2 = \mathcal{P}$. (It need not be Hermitian with respect to any scalar product.) If $\mathcal{P}$ is a projector then so is its complement $\mathcal{Q} := 1 - \mathcal{P}$. For our purposes we consider projectors which project arbitrary vectors in Liouville space onto the level of description; i.e., for which

$$\mathcal{P} A = A \quad \text{iff} \quad A \in \text{span}\{1, G_a\} \quad . \tag{34}$$

The relevant observables and therefore the level of description are assumed to have no explicit time-dependence. In contrast, one allows the projector to depend on the expectation values $\{g_a(t)\}$ of the relevant observables, thus making it an implicit function of time: $\mathcal{P} \equiv \mathcal{P}[g_a(t)]$ (cf. Fig. 1). Since the level of description itself remains fixed, a second projection will have no effect, even at a different time:

$$\mathcal{P}[g_a(t_2)] \mathcal{P}[g_a(t_1)] = \mathcal{P}[g_a(t_1)] \quad \forall \, t_1, t_2 \quad . \tag{35}$$

**Two important examples are the Kawasaki-Gunton and the Mori projector.** In principle, there are many possible choices for the projector, but the following two are particularly useful:

– The *Kawasaki-Gunton projector* [63] is defined such that the projection onto the level of description is orthogonal with respect to the canonical correlation function $\langle \cdot ; \cdot \rangle_{\text{rel}(t)}$. The correlation function is defined in the time-dependent *relevant part of the state*, $\rho_{\text{rel}}(t)$. The Kawasaki-Gunton projector thus varies in time: it depends –via $\rho_{\text{rel}}(t)$– on the time-dependent expectation values $\{g_a(t)\}$. Explicitly it reads

$$\mathcal{P}_{\text{KG}}[g_a(t)]\, A := \langle 1; A\rangle_{\text{rel}(t)} 1 + (C_{\text{rel}(t)}^{-1})^{ba} \langle \delta G_a^{\text{rel}(t)}; A\rangle_{\text{rel}(t)} \delta G_b^{\text{rel}(t)} \quad . \tag{36}$$

Here the operators $\delta G_a^{\text{rel}(t)}$ are defined as

$$\delta G_a^{\text{rel}(t)} := G_a - g_a(t) \quad , \tag{37}$$

and $C_{\text{rel}(t)}$ denotes the correlation matrix evaluated in $\rho_{\text{rel}}(t)$.

– The *Mori projector* [7] is defined such that the projection onto the level of description is orthogonal with respect to the canonical correlation function $\langle \cdot; \cdot \rangle_{\text{eq}}$. The correlation function is now defined in *equilibrium*, $\rho_{\text{eq}}$. The Mori projector is thus time-independent. Explicitly it reads

$$\mathcal{P}_{\text{M}} A := \langle 1; A\rangle_{\text{eq}} 1 + (C^{-1})^{ba} \langle \delta G_a; A\rangle_{\text{eq}} \delta G_b \quad . \tag{38}$$

Here the operators $\delta G_a$ are defined as

$$\delta G_a := G_a - \langle G_a\rangle_{\text{eq}} \quad , \tag{39}$$

and $C$ denotes the correlation matrix evaluated in equilibrium.

The above two projectors ('KG' and 'M') lead to different expectation values of the projected observable:

$$(\rho(t)|\mathcal{P} A) = \begin{cases} \langle A\rangle_{\text{rel}(t)} & : \text{KG} \\ \langle A\rangle_{\text{eq}} + (C^{-1})^{ba} \langle \delta G_a; A\rangle_{\text{eq}} \cdot [g_b(t) - \langle G_b\rangle_{\text{eq}}] & : \text{M} \end{cases} \tag{40}$$

Provided the equilibrium state is completely characterized by the expectation values $\{\langle G_a\rangle_{\text{eq}}\}$, i. e., provided the relevant observables contain all constants of the motion as a subset, then according to (33) the two expectation values agree to first order in $[g_b(t) - \langle G_b\rangle_{\text{eq}}]$. In other words, in the linear regime close to equilibrium the Kawasaki-Gunton and Mori projectors yield identical results. Far from equilibrium, however, the two projections may have completely different effects. Which of the two projectors is better suited to a particular application differs from case to case; their respective advantages and disadvantages will be discussed in section 3.4.2.

*3.2.3 Equation of Motion for the Relevant Observables*

**Eliminating irrelevant degrees of freedom gives rise to a memory term and a residual force term.** We return to the generic case of an arbitrary, possibly time-dependent projector $\mathcal{P}(t) \equiv \mathcal{P}[g_a(t)]$. The only specific property that we require is that it satisfy

$$(\rho(t)|\frac{\mathrm{d}}{\mathrm{d}t}\mathcal{P}(t)\,A) = 0 \quad \forall\,A \quad . \tag{41}$$

Both the Kawasaki-Gunton and the Mori projector fulfill this requirement. As before, we assume that the Hamiltonian and hence the Liouvillian are not explicitly time-dependent.

While $\mathcal{U}(t_0, t)$ is the evolution superoperator corresponding to the Liouvillian $\mathcal{L}$, let $\mathcal{T}(t_0, t)$ be the evolution superoperator corresponding to

$$\mathcal{L}_{\mathrm{irr}}(t) := \mathcal{Q}(t)\mathcal{L}\mathcal{Q}(t) \quad ; \tag{42}$$

i. e., it satisfies the equation of motion

$$\frac{\partial}{\partial t}\mathcal{T}(t_0, t) = i\,\mathcal{T}(t_0, t)\mathcal{L}_{\mathrm{irr}}(t) \tag{43}$$

or

$$\frac{\partial}{\partial t'}\mathcal{T}(t', t) = -i\,\mathcal{L}_{\mathrm{irr}}(t')\mathcal{T}(t', t) \quad , \tag{44}$$

respectively, with initial condition $\mathcal{T}(t_0, t_0) = 1$. Since $\mathcal{Q}$ projects out the irrelevant component of an observable, the superoperators $\mathcal{L}_{\mathrm{irr}}$ and hence $\mathcal{T}$ may be pictured as describing the evolution of the *irrelevant* degrees of freedom of the system. Using the definition (44) of $\mathcal{T}$, the Liouville-von Neumann equation (9), as well as the property (41) of projectors, we find

$$\begin{aligned}(\rho(t)|\mathcal{Q}(t) &= (\rho(t_0)|\mathcal{Q}(t_0)\mathcal{T}(t_0, t) \\ &+ \int_{t_0}^{t} \mathrm{d}t' \left[\left(\frac{\mathrm{d}}{\mathrm{d}t'}(\rho(t')|\mathcal{Q}(t'))\right)\mathcal{T}(t', t) + (\rho(t')|\mathcal{Q}(t')\frac{\partial}{\partial t'}\mathcal{T}(t', t)\right] \\ &= (\rho(t_0)|\mathcal{Q}(t_0)\mathcal{T}(t_0, t) + i\int_{t_0}^{t} \mathrm{d}t'\,(\rho(t')|\mathcal{P}(t')\mathcal{L}\mathcal{Q}(t')\mathcal{T}(t', t) \quad . \end{aligned} \tag{45}$$

Without changing the left-hand side we may multiply this equation by $\mathcal{Q}(t)$ from the right. The equation of motion for the expectation values of the relevant observables can then be written as

$$\dot{g}_a(t) = \frac{\mathrm{d}}{\mathrm{d}t}(\rho(t)|G_a)$$

$$\begin{aligned}
&= i(\rho(t)|\mathcal{P}(t)\mathcal{L}G_a) + i(\rho(t)|\mathcal{Q}(t)\mathcal{L}G_a) \\
&= i(\rho(t)|\mathcal{P}(t)\mathcal{L}G_a) - \int_{t_0}^{t} dt'\, (\rho(t')|\mathcal{P}(t')\mathcal{L}\mathcal{Q}(t')\mathcal{T}(t',t)\mathcal{Q}(t)\mathcal{L}G_a) \\
&\quad + i(\rho(t_0)|\mathcal{Q}(t_0)\mathcal{T}(t_0,t)\mathcal{Q}(t)\mathcal{L}G_a) \quad .
\end{aligned} \tag{46}$$

This is the desired result. The equation of motion is exact: no physical assumptions or approximations have been made. We have largely succeeded in replacing $(\rho| \to (\rho|\mathcal{P}$, thus eliminating almost completely the influence of irrelevant degrees of freedom. The only exception is the last term which accounts for the residual influence of irrelevant components in the initial state; but in many applications this term can be shown to vanish, or else to be manageable.

Of course, the elimination of irrelevant degrees of freedom has its price. Comparing the above equation of motion with the Liouville-von Neumann equation, we notice that it contains two terms which are qualitatively new: (i) an integral ("memory") term, containing contributions from all times between the initial and the present time; and (ii) a "residual force" term describing the effect of irrelevant components in the initial state. The physical meaning of these two terms can be easily discerned if read from left to right: (i) At time $t' < t$ relevant degrees of freedom (projected out by $\mathcal{P}$) couple via an interaction ($\mathcal{L}$) to irrelevant degrees of freedom (projected out by $\mathcal{Q}$), which subsequently evolve in time ($\mathcal{T}$) and, due to a second interaction ($\mathcal{L}$), acquire relevancy again, thus influencing the evolution of the relevant observable $G_a$ at the present time $t$. (ii) Irrelevant components in the initial state ($\mathcal{Q}$) evolve in time ($\mathcal{T}$) and, due to interaction ($\mathcal{L}$), acquire relevancy at the present time $t$.

**The projected equation of motion is a good starting point for approximations.** In essence, the influence of irrelevant degrees of freedom has been mapped onto a non-locality in time: future expectation values of the relevant observables are predicted not just on the basis of their present values, but based on their entire history. At first sight such an equation of motion may seem even more complicated and awkward than the original Liouville-von Neumann equation; and indeed, in many cases it is just as hard to solve *exactly*. However, mapping the complexity of the problem onto a non-local behavior in time opens the way to the systematic exploitation of a possible separation of time scales. Therefore, the above equation of motion serves as a much better starting point for various approximations, in particular the Markovian approximation. In fact, many of the well-known equations of non-equilibrium statistical mechanics can be derived from there, either as special cases or with the help of a Markovian approximation. Examples that we will discuss in detail are the Robertson and Langevin-Mori equations and, later on, the rate equation for elastic scattering and the quantum Boltzmann equa-tion; other examples being the Nakajima-Zwanzig, Master, or time-dependent Hartree-Fock equations [17].

### 3.2.4 Approximations

**Two important approximations are perturbation theory and the Markovian limit.** The projection method was initially motivated by the claim that it permits a new kind of approximation based on the existence of disparate time scales in the system: the Markovian approximation. The general equation of motion for the relevant observables thus serves as the starting point for *two* important approximations:

– the conventional perturbation expansion; and
– the Markovian limit, as well as (associated with it) the quasistationary limit.

These two approximations will be discussed separately. In order to streamline notation, we will denote the kernel of the memory term by

$$\mathcal{K}(t',t) := \mathcal{P}(t')\mathcal{L}\mathcal{Q}(t')\mathcal{T}(t',t)\mathcal{Q}(t)\mathcal{L}\mathcal{P}(t) \quad . \tag{47}$$

**Perturbation theory is based on a suitable decomposition of the Liouvillian.** Often the Liouvillian can be split into a free part and an interaction part,

$$\mathcal{L} = \mathcal{L}^{(0)} + \mathcal{V} \quad , \tag{48}$$

corresponding to a decomposition $H = H^{(0)} + V$ of the Hamiltonian. Provided free evolution does not mix relevant and irrelevant degrees of freedom, i.e., provided

$$[\mathcal{L}^{(0)}, \mathcal{P}(t)] = 0 \quad , \tag{49}$$

then in the memory term $\mathcal{P}\mathcal{L}\mathcal{Q} = \mathcal{P}\mathcal{V}\mathcal{Q}$ and $\mathcal{Q}\mathcal{L}\mathcal{P} = \mathcal{Q}\mathcal{V}\mathcal{P}$. In that case the memory term is at least of order $O(\mathcal{V}^2)$; it can be dropped in first order perturbation theory. One prominent application of first order perturbation theory is to a system of nonrelativistic fermions with effective 4-fermion interaction: if one chooses the level of description to be spanned by all single-particle observables and employs the Kawasaki-Gunton projector, then dropping the memory term will yield the time-dependent Hartree-Fock equations [64,65]. In this example first order perturbation theory amounts to a mean field approximation.

In order to obtain higher order contributions one must include the memory term and expand

$$\mathcal{Q}(t')\mathcal{T}(t',t)\mathcal{Q}(t) = \mathcal{Q}(t')\mathcal{T}^{(0)}(t',t)\mathcal{Q}(t) + O(\mathcal{V}) \quad , \tag{50}$$

where $\mathcal{T}^{(0)}$ is a time-ordered exponential of $\mathcal{L}_{\text{irr}}^{(0)}(t)$. Using $\mathcal{Q}(t_2)\mathcal{Q}(t_1) = \mathcal{Q}(t_2)$, all the $\mathcal{Q}$'s appearing in $\mathcal{T}^{(0)}$ can be shuffled to the left and absorbed into $\mathcal{Q}(t')$, allowing one to replace $\mathcal{T}^{(0)}$ simply by $\mathcal{U}^{(0)}$. To second order we thus obtain

$$\begin{aligned}\dot{g}_a^{(2)}(t) =\;& i(\rho(t)|\mathcal{P}(t)\mathcal{L}G_a) \\ & - \int_0^{t-t_0} d\tau\, (\rho(t-\tau)|\mathcal{P}(t-\tau)\mathcal{V}\mathcal{Q}(t-\tau)\mathcal{U}^{(0)}(0,\tau)\mathcal{Q}(t)\mathcal{V}G_a) \\ & + \text{residual force} \quad . \end{aligned} \tag{51}$$

**The Markovian limit amounts to ignoring the system's past history.** We have found that predictions of future expectation values of the relevant observables generally depend in a complicated manner on both their present values and their past history. There are thus two distinct time scales: (i) the scale $\tau_{\text{rel}}$ –or several scales $\{\tau_{\text{rel}}^{(i)}\}$– on which the expectation values $\{g_a(t)\}$ evolve; and (ii) the *memory time* $\tau_{\text{mem}}$ which characterizes the length of the time interval that contributes significantly to the memory term. Loosely speaking, the memory time determines how far back into the past one has to reach in order to make predictions for the further evolution of the relevant observables. Often the memory kernel $\mathcal{K}(t',t)$ is proportional to some distribution (e. g., Gaussian or Lorentzian) in $(t - t')$, in which case the memory time can be identified with the width of that distribution. Symbolically:

$$\mathcal{K}(t',t) \approx 0 \quad \text{if} \quad (t - t') > \tau_{\text{mem}} \quad . \tag{52}$$

If this memory time is small compared to the typical time scale on which the relevant observables evolve, $\tau_{\text{mem}} \ll \tau_{\text{rel}}$, then memory effects can be neglected and predictions for the relevant observables be based solely on their present values. One may then assume that in the memory term $g_a(t') \approx g_a(t)$, and hence replace

$$\begin{aligned}\mathcal{P}[g_a(t')] &\to \mathcal{P}[g_a(t)] \\ (\rho(t')|\mathcal{P}(t') &\to (\rho(t)|\mathcal{P}(t) \\ \mathcal{T}(t',t) &\to \exp[i\mathcal{Q}(t)\mathcal{L}\mathcal{Q}(t)\cdot(t-t')] \quad . \end{aligned} \tag{53}$$

With these replacements, the equation of motion –ignoring the residual force term– simplifies to

$$\dot{g}_a{}^{(\mathrm{m})}(t) = i(\rho(t)|\mathcal{P}(t)\mathcal{L}G_a)$$
$$-(\rho(t)|\mathcal{P}(t)\mathcal{L}\mathcal{Q}(t)\int_0^{t-t_0} \mathrm{d}\tau\ \exp[i\mathcal{Q}(t)\mathcal{L}\mathcal{Q}(t)\cdot\tau]\,\mathcal{Q}(t)\mathcal{L}G_a) \quad . \quad (54)$$

The change of expectation values of the relevant observables now depends only on their present values, their past history being completely ignored. This is the *Markovian limit*. Note that nevertheless the kernel $\mathcal{K}$ still contributes: ignoring the system's past history does not mean that the memory term is eliminated altogether. If on a given level of description it is justified to take this Markovian limit, the level is called a *Markovian level of description*.

Closely related to the Markovian limit is the *quasistationary limit*. For times $t$ with $(t - t_0) \gg \tau_{\mathrm{mem}}$ the exact "location" of the initial time $t_0$ becomes irrelevant; one may then let $t_0 \to -\infty$. This removal of the cut-off $t_0$ restores temporal homogeneity and hence energy conservation.

**One may systematically expand around the Markovian limit.** Expanding in the memory term $g_a(t')$ around $g_a(t)$,

$$g_a(t') = g_a(t) + \dot{g}_a(t)\cdot(t-t') + \ldots \quad , \quad (55)$$

will yield corrections to the Markovian equation of motion. These non-Markovian corrections are of successively higher order in $(\tau_{\mathrm{mem}}/\tau_{\mathrm{rel}})$. To illustrate this, we consider a simple model: the evolution of a single expectation value $g(t)$ governed by a non-Markovian equation of motion

$$\dot{g}(t) = -\int_0^{t-t_0} \mathrm{d}\tau\ \gamma(\tau) g(t-\tau) \quad . \quad (56)$$

For simplicity we assume the kernel $\gamma(\tau)$ to have the form

$$\gamma(\tau) = \frac{\Gamma}{\tau_{\mathrm{mem}}}\exp(-\tau/\tau_{\mathrm{mem}}) \quad (57)$$

with a finite memory time $\tau_{\mathrm{mem}}$. As long as $\tau_{\mathrm{mem}} \ll \Gamma^{-1}$, it is justified to take the Markovian limit; one may then replace $g(t-\tau) \to g(t)$ and obtain the approximate equation of motion

$$\dot{g}^{(\mathrm{m})}(t) = -\Gamma g(t) \quad . \quad (58)$$

If one wants to go beyond this approximation one must expand

$$g(t-\tau) = g(t) + \sum_{k=1}^{\infty} \frac{1}{k!}(-\tau)^k \frac{\mathrm{d}^k}{\mathrm{d}t^k} g(t) \quad, \tag{59}$$

which yields the *exact* equation

$$\dot{g}(t) = \dot{g}^{(\mathrm{m})}(t) - \Gamma \sum_{k=1}^{\infty} (-\tau_{\mathrm{mem}})^k \frac{\mathrm{d}^k}{\mathrm{d}t^k} g(t) \quad. \tag{60}$$

Non-Markovian corrections are thus associated with successively higher ($k$-th) order derivatives of $g(t)$. Provided the expectation value evolves on a typical time scale $\tau_{\mathrm{rel}}$, it is

$$\frac{1}{g(t)} \frac{\mathrm{d}^k}{\mathrm{d}t^k} g(t) \sim O(1/\tau_{\mathrm{rel}}^k) \quad, \tag{61}$$

and hence the $k$-th non-Markovian correction is of order

$$\frac{1}{\dot{g}^{(\mathrm{m})}(t)} \left[ \Gamma \tau_{\mathrm{mem}}^k \frac{\mathrm{d}^k}{\mathrm{d}t^k} g(t) \right] \sim O((\tau_{\mathrm{mem}}/\tau_{\mathrm{rel}})^k) \quad. \tag{62}$$

*3.3  Which observables should be relevant?*

**The proper choice of relevant observables is dictated by the time scales.** In principle, the choice of the set of "relevant observables" is arbitrary. However, some choices are more useful than others. At first sight, it seems obvious that always those observables should be considered relevant which are actually being monitored in the experiment; but this choice is not always the most suitable. As an example consider an experiment performed on a dilute classical gas, in the course of which local energy and particle densities are being monitored. A theoretical description of this experiment will most likely start from the Boltzmann equation — even though in the Boltzmann equation *all* single-particle observables are taken to be relevant, not just those which are actually being monitored. The reason for employing the Boltzmann equation, rather than an equation of motion for the local densities only, lies in the fact that the level of description spanned by *all* single-particle observables is Markovian, while the smaller experimental level is not.

Such a situation is quite common. Typically, the experimentally monitored observables are contained in some larger set of slowly varying observables. (For an observable to be measurable in practice it is usually necessary that it vary slowly.) While the experimental level of description is not Markovian, the level associated with the larger set is. In this case it is best to *extend* the level of description from the experimental to this larger Markovian level. Only this allows one to make use of the powerful Markovian approximation, a significant advantage in any explicit calculation. The proper choice of the relevant observables is thus dictated by the time scales.

## 3.4 Representations of the Equation of Motion

Representations of the general equation of motion can be obtained by choosing specific projectors $\mathcal{P}(t)$. We will discuss two choices: the Kawasaki-Gunton projector, which leads to the *Robertson equation* [9–12,63]; and the Mori projector, which leads to the *Langevin-Mori equation* [7,8]. In both cases we will obtain the respective representation by inserting Eq. (40) into Eq. (46).

### 3.4.1 Robertson and Langevin-Mori Equations

**Inserting the Kawasaki-Gunton projector into the general equation of motion yields the Robertson equation:**

$$\dot{g}_a(t) = i\langle \mathcal{L}G_a \rangle_{\mathrm{rel}(t)} - \int_{t_0}^{t} \mathrm{d}t' \, \langle \mathcal{L}\mathcal{Q}(t')\mathcal{T}(t',t)\mathcal{Q}(t)\mathcal{L}G_a \rangle_{\mathrm{rel}(t')}$$
$$+ \langle \mathcal{T}(t_0,t)\mathcal{Q}(t)i\mathcal{L}G_a \rangle_{\mathrm{irr}(t_0)} \quad . \tag{63}$$

Here

$$\rho_{\mathrm{irr}}(t_0) := \rho(t_0) - \rho_{\mathrm{rel}}(t_0) \tag{64}$$

denotes the *irrelevant part* of the initial state. With the help of the identity

$$\langle i\mathcal{L}A \rangle_{\mathrm{rel}(t)} = \langle i\mathcal{L}G_a; A \rangle_{\mathrm{rel}(t)} \lambda^a(t) \quad \forall A \quad , \tag{65}$$

and using the fact that the projector $\mathcal{Q}(t')$ is Hermitian with respect to $\langle \cdot; \cdot \rangle_{\mathrm{rel}(t')}$, one may also write

$$\dot{g}_a(t) = \langle i\mathcal{L}G_c; G_a \rangle_{\mathrm{rel}(t)} \lambda^c(t) + \int_{t_0}^{t} \mathrm{d}t' \, \langle \mathcal{Q}(t')i\mathcal{L}G_c; \mathcal{T}(t',t)\mathcal{Q}(t)i\mathcal{L}G_a \rangle_{\mathrm{rel}(t')} \lambda^c(t')$$
$$+ \langle \mathcal{T}(t_0,t)\mathcal{Q}(t)i\mathcal{L}G_a \rangle_{\mathrm{irr}(t_0)} \quad . \tag{66}$$

Inserting the Mori projector into the general equation of motion yields the Langevin-Mori equation:

$$\dot{g}_a(t) = \Omega_a^b \delta g_b(t) - \int_0^{t-t_0} d\tau \, \gamma_a^b(\tau) \delta g_b(t - \tau) + \langle \mathcal{T}(t_0, t) f_a \rangle_{\rho(t_0)} \quad . \tag{67}$$

Here

$$\Omega_a^b := -(C^{-1})^{bc} \langle i\mathcal{L}G_c; G_a \rangle_{\text{eq}} \tag{68}$$

denotes the *frequency matrix*;

$$\gamma_a^b(\tau) := (C^{-1})^{bc} \langle f_c; \mathcal{T}(0, \tau) f_a \rangle_{\text{eq}} \tag{69}$$

the *memory matrix*;

$$f_a := \mathcal{Q} i\mathcal{L} G_a \tag{70}$$

the *stochastic force*; and finally $\delta g_a(t)$ is defined as

$$\delta g_a(t) := g_a(t) - \langle G_a \rangle_{\text{eq}} \quad . \tag{71}$$

The Langevin-Mori equation and the second version (66) of the Robertson equation have similar structures. The former can be obtained from the latter by replacing

$$\langle \cdot; \cdot \rangle_{\text{rel}(t)} \to \langle \cdot; \cdot \rangle_{\text{eq}} \; , \quad \mathcal{Q}(t) i\mathcal{L} G_a \to f_a \; , \quad \lambda^c(t) \to -(C^{-1})^{bc} \delta g_b(t) \quad . \tag{72}$$

**Langevin-Mori theory leads to an equation of motion for the dynamical correlations.** The Langevin-Mori equation may also be read as an operator equation if one replaces the expectation values $\delta g_a(t)$ by Heisenberg-picture operators $\delta G_a(t)$. The resulting operator equation can be inserted into the definition of the *normalized dynamical correlation matrix* [17]

$$\Xi_a^b(t) := (C^{-1})^{bc} \langle \delta G_c; \delta G_a(t) \rangle_{\text{eq}} \quad , \tag{73}$$

with $\Xi_a^b(0) = \delta_a^b$, to obtain its equation of motion

$$\dot{\Xi}(t) = \Omega \Xi(t) - \int_0^t d\tau \, \gamma(\tau) \Xi(t - \tau) \quad . \tag{74}$$

Here we have employed a compact matrix notation. Note that both the initial time $t_0$ and the residual force term have dropped out.

Multiplying $\Xi(t)$ and $\gamma(t)$ by step functions yields the *causal* normalized dynamical correlation matrix $\Xi_<(t)$ and *causal* memory matrix $\gamma_<(t)$, respectively. Their Fourier transforms are given by

$$\Xi^b_{<a}(\omega) = (C^{-1})^{bc} \langle \delta G_c ; \mathcal{G}_<(\omega) \delta G_a \rangle_{\text{eq}} \quad , \tag{75}$$

$$\gamma^b_{<a}(\omega) = (C^{-1})^{bc} \langle f_c ; \mathcal{T}_<(\omega) f_a \rangle_{\text{eq}} \quad . \tag{76}$$

As a consequence of the equation of motion (74), the Fourier transformed $\Xi_<(\omega)$ can be expressed directly in terms of the frequency and memory matrices:

$$\Xi_<(\omega) = [\gamma_<(\omega) - \Omega - i\omega \cdot 1]^{-1} \quad . \tag{77}$$

In the Markovian limit the memory matrix falls off rapidly in time, hence its Fourier transform is practically constant; one may then replace $\gamma_<(\omega)$ with $\gamma_<(0)$.

*3.4.2 Which representation should one choose?*

**Whether the Kawasaki-Gunton or the Mori projector is more advantageous depends on the specific application.** Since the Robertson equation and the Langevin-Mori equation are both exact –although in practice they can only be solved with suitable approximations– the choice of projector is not a matter of principle but of convenience. Depending upon the specific physical process under consideration, either the Kawasaki-Gunton projector and hence the Robertson equation, or the Mori projector and hence the Langevin-Mori equation, may serve as better starting points for explicit calculations.

The Kawasaki-Gunton projector has one major disadvantage:

– The Robertson equation leads to a system of coupled integro-differential equations for the expectation values $\{g_a(t)\}$. Since the Kawasaki-Gunton projector is not constant in time, but depends in a non-trivial way on the $\{g_a(t)\}$, these equations are generally nonlinear.

On the other hand, employing the Kawasaki-Gunton projector has three potential advantages:

– Often the initial macrostate can be characterized completely by the expectation values $\{\langle G_a \rangle(t_0)\}$ of the relevant observables. The initial state is then determined by maximizing its von Neumann entropy, and hence has just the generalized canonical form. In that case the initial state has no irrelevant component, which in turn implies that there is no residual force. Eliminating the residual force term in this manner greatly simplifies the calculation.

– If all relevant observables commute, i.e., if $[G_a, G_b] = 0$, then also the first term of the equation of motion vanishes:

$$(\rho_{\rm rel}|\mathcal{L}G_a) \propto {\rm tr}(\exp(-\lambda^b G_b)[H, G_a]) = {\rm tr}([G_a, \exp(-\lambda^b G_b)]H) = 0 \,. (78)$$

This in turn implies that $\mathcal{PLP} = 0$ and that hence in the memory term of the Robertson equation one may omit all complementary projectors $\mathcal{Q}$. However, these conclusions must be taken with a grain of salt: their derivation depends on the cyclic permutability of operators under the trace, a property which holds only if both $G_a$ and $H$ are bounded [66]. Indeed, in section 4.2 we will encounter an example –the acceleration term in the quantum Boltzmann equation– where the above line of reasoning does not apply.

– Many transport theories have as their relevant observables some set of number operators $\{N_i\}$ or, more generally, of single-particle observables $\{a_i^\dagger a_k\}$. Evaluation of the memory term then requires calculating expectation values $\langle a_i^\dagger a_j a_k^\dagger a_l \ldots \rangle_{\rm rel(t')}$ of products of several field operators. Because the relevant part $\rho_{\rm rel}(t')$ has the so-called Hartree form [67] (cf. table 1), Wick's theorem can be applied and these expectation values be evaluated.

The advantages and disadvantages of the Kawasaki-Gunton projector are in a sense complementary to those of the Mori projector. The main disadvantage of the latter is:

– The residual force term generally does not vanish, nor can it be calculated exactly. Rather, in most applications it must be suitably modelled.

This has to be weighed against its advantages:

– Due to the time-independence of the Mori projector, the Langevin-Mori equation has a convenient linear structure. This linearity permits Fourier transformation and thus a treatment of the dynamics in the frequency representation.
– Frequency and memory matrices are defined in equilibrium and may therefore be easier to evaluate.
– In the context of linear response theory, evaluation of the Kubo formula can often be reduced to calculating the dynamical correlation matrix $\Xi_<(\omega)$ of some selected observables. This dynamical correlation matrix, in turn, is most easily calculated in the Mori formalism (Eq. (77)). There the problematic residual force term no longer appears.

*Summary.* To put it in a nutshell, there is usually a trade-off: complicated features of the dynamics may appear either in the form of nonlinearities (Kawasaki-Gunton) or a non-vanishing residual force (Mori). Which projector one should choose depends on which complication is easier to handle. As a rule of thumb, the Kawasaki-Gunton projector is usually better suited for the description of processes far from equilibrium, while the Mori projector is better adapted to describing processes in the linear regime close to equilibrium, and to evaluating the Kubo formula.

3.5  *Accompanying Entropy*

**Responsible for entropy generation is the memory term.** Associated with the relevant part of the statistical operator is the *relevant entropy*

$$S_{\rm rel}[g_a(t)] := -k \, {\rm tr}(\rho_{\rm rel}(t) \ln \rho_{\rm rel}(t)) \quad . \tag{79}$$

This relevant entropy is also known as the *accompanying entropy with respect to the level of description* [17]. Its value generally varies in time. According to Eq. (28), relevant entropy is produced at the rate

$$\dot{S}_{\rm rel}[g_a(t)] = k \, \lambda^a(t) \dot{g}_a(t) \quad . \tag{80}$$

This formula can be evaluated explicitly by inserting the Robertson equation (66). We assume that the initial state is completely specified by the initial expectation values of the relevant observables so that there is no residual force. One can show that the first term in the Robertson equation never contributes to entropy generation; hence the only non-vanishing contribution stems from the memory term:

$$\dot{S}_{\rm rel}[g_a(t)] = k \int_{t_0}^{t} dt' \, \langle \lambda^c(t') \mathcal{Q}(t') i\mathcal{L} G_c; \mathcal{T}(t',t) \lambda^a(t) \mathcal{Q}(t) i\mathcal{L} G_a \rangle_{{\rm rel}(t')} \quad . \tag{81}$$

This expression is valid arbitrarily far from equilibrium.

**In the Markovian limit the accompanying entropy increases monotonically.** This statement is true generally; however, for simplicity, we will prove it here in the linear regime close to equilibrium. In that regime the above expression for the entropy production rate can be further simplified: one may replace Kawasaki-Gunton by Mori projectors and evaluate the canonical correlation function in equilibrium. Taking the quasistationary limit ($t_0 \to -\infty$), the entropy production rate then reads

$$\dot{S}_{\text{rel}}^{\text{lin}}[g_a(t)] = k \int_{-\infty}^{\infty} d\tau\, \lambda^c(t-\tau)\langle f_c; \mathcal{T}_<(0,\tau)f_a\rangle_{\text{eq}}\lambda^a(t) \quad . \tag{82}$$

Provided the evolution is Markovian, this reduces to

$$\dot{S}_{\text{rel}}^{\text{lin}}[g_a(t)] = \pi k \langle \lambda^c(t)f_c; \delta(\mathcal{QLQ})\,\lambda^a(t)f_a\rangle_{\text{eq}} \geq 0 \quad . \tag{83}$$

The superoperator $\delta(\mathcal{QLQ})$ and therefore the entropy production rate are non-negative, Q.E.D.

This result is just one particular manifestation of the far more general $H$-theorem. The $H$-theorem, the second law of thermodynamics, and their conceptual ramifications will be discussed in greater detail in the appendix.

# 4 Application: Quantum Transport Equations

## 4.1 Introduction

**When applying the projection method one generally proceeds in eight steps:**

 (i) Choose the level of description by identifying the relevant observables.
 (ii) Determine, or make assumptions about, the initial state.
 (iii) Determine the microscopic dynamics (Hamiltonian).
 (iv) Choose the projector and hence the representation of the equation of motion.
 (v) Inserting the specific choice of relevant observables, initial state and Hamiltonian, obtain the exact, generally *non*-Markovian equation of motion.
 (vi) First part of the time scale analysis: Determine the memory time.
 (vii) Second part of the time scale analysis: Take the Markovian limit and determine the typical time scale on which the relevant observables evolve; then, by comparing this time scale with the memory time, determine *a posteriori* whether the Markovian limit has been justified. In this manner, obtain physical criteria for the validity of the Markovian limit.
 (viii) Provided that the Markovian approximation *is* justified in the physical regime of interest, recover a Markovian equation of motion. If not,
   – solve a non-Markovian equation of motion; or
   – systematically expand around the Markovian limit; or
   – extend the level of description, then repeat the time scale analysis.

**Most quantum transport equations, notably rate equations and the quantum Boltzmann equation, refer to the evolution of single-particle distributions.** Relevant are thus certain number operators

$$\{N_i\} := \{a_i^\dagger a_i\} \quad . \tag{84}$$

Associated with this single-particle level of description is a relevant part of the statistical operator of the form

$$\rho_{\rm rel}(t) = Z(t)^{-1} \exp(-\lambda^i(t) N_i) \quad , \tag{85}$$

the Lagrange parameters being adjusted so as to yield, at each time $t$, the correct occupation numbers

$$n_i(t) := \langle N_i \rangle(t) \quad . \tag{86}$$

It is commonly assumed that the states $\{|i\rangle\}$ constitute a complete orthonormal basis; hence the observable

$$N := \sum_i N_i \tag{87}$$

represents the total number of particles. Furthermore, it is assumed that there are "no initial correlations," i. e., that

$$\rho(t_0) = \rho_{\text{rel}}(t_0) \quad . \tag{88}$$

The resulting transport equation has the structure

$$\dot{n}_i(t) = F[\{n_j(t')|t_0 \leq t' \leq t\}] \quad , \tag{89}$$

which only in the Markovian limit simplifies to

$$\dot{n}_i^{(m)}(t) = F^{(m)}[\{n_j(t)\}] \quad . \tag{90}$$

Such a Markovian limit is often tacitly assumed without rigorous justification. Here we wish to proceed more carefully: we first employ the projection method to derive the exact *non*-Markovian equation of motion; then we perform a thorough time scale analysis; obtain criteria for the validity of the Markovian limit; and only where appropriate, recover the usual Markovian expression.

Different quantum transport equations, or individual terms therein, differ from each other in the choice of basis states $\{|i\rangle\}$ or in the form of the underlying microscopic dynamics. In the following sections we will discuss four specific microscopic processes –acceleration, level transitions, spontaneous pair creation and binary collisions– and investigate how they manifest themselves in quantum transport equations. The examples are summarized in Table 2.

**For later reference we collect some useful formulae.** The field operators satisfy commutation or anticommutation relations,

$$[a_i, a_j]_\mp = [a_i^\dagger, a_j^\dagger]_\mp = 0 \quad , \tag{91}$$

$$[a_i, a_j^\dagger]_\mp = \delta_{ij} \quad , \tag{92}$$

depending on whether they describe bosons (upper sign) or fermions (lower sign). Since the relevant part of the statistical operator has the so-called

Table 2
Four examples of quantum transport equations or parts thereof. The table lists the underlying microscopic dynamics, the choice of basis states ($|\mathbf{p}\rangle$=momentum eigenstates, $|\epsilon_0^i\rangle$=eigenstates of the unperturbed Hamiltonian), the choice of the projector (M=Mori, KG=Kawasaki-Gunton), the interesting term in the general equation of motion (f=first term, m=memory term), and the resulting transport equation or part thereof (QBE=quantum Boltzmann equation).

| Microscopic dynamics | $|i\rangle$ | $\mathcal{P}$ | term | gives rise to |
|---|---|---|---|---|
| acceleration in electric field | $|\mathbf{p}\rangle$ | — | f | acceleration term in QBE |
| level transitions due to perturbing potential | $|\epsilon_0^i\rangle$ | M/KG | m | rate equation for elastic scattering |
| spontaneous pair creation in electric field | $|\mathbf{p}\rangle$ | M/KG | m | source term in QBE |
| binary collisions | $|\mathbf{p}\rangle$ | KG | m | collision term in QBE |

Hartree form (cf. table 1), arbitrary expectation values of products of field operators can be calculated with the help of Wick's theorem [67]:

$$\langle ABC \ldots F \rangle_{\mathrm{rel}} = [A^\bullet B^\bullet C^{\bullet\bullet} \ldots F^{\bullet\bullet\bullet}] + [A^\bullet B^{\bullet\bullet} C^\bullet \ldots F^{\bullet\bullet\bullet}] + \ldots \quad . \tag{93}$$

Here we have employed several shorthand notations. $A, B, \ldots$ represent arbitrary annihilation or creation operators; and the "bullets" $\bullet$ indicate *contractions*, defined by

$$A^\bullet B^\bullet := \langle AB \rangle_{\mathrm{rel}} \quad . \tag{94}$$

Furthermore, it is understood that, e.g.,

$$[A^\bullet B^{\bullet\bullet} C^\bullet \ldots F^{\bullet\bullet\bullet}] = \pm [A^\bullet C^\bullet B^{\bullet\bullet} \ldots F^{\bullet\bullet\bullet}] \quad , \tag{95}$$

the sign depending on whether the particles under consideration are bosons or fermions. More generally, each rearrangement $r$ of the field operators must be accompanied by a factor

$$\varepsilon(r) := (\pm 1)^{\deg(r)} \tag{96}$$

which is determined by the degree of the permutation involved. Note that not all permutations are permitted: the order of two operators which belong to the same contraction must be maintained. Wick's theorem can then be written in the symbolic form

$$\langle \prod \ldots \rangle_{\mathrm{rel}} = \sum_r \varepsilon(r) \prod \text{contractions} \quad . \tag{97}$$

Each individual contraction may be evaluated using

$$\langle a_i^\dagger a_j \rangle_{\rm rel} = \delta_{ij} n_i = \delta_{ij}[e^{\lambda^i} \mp 1]^{-1} \quad . \tag{98}$$

We also note the correlation matrix

$$C_{ij} := \langle \delta N_i ; \delta N_j \rangle_{\rm rel} = \delta_{ij} n_i (1 \pm n_i) \quad . \tag{99}$$

4.2 Acceleration Term

**Acceleration gives rise to a non-vanishing first term in the general equation of motion.** Our first example is almost trivial. We consider the acceleration of charged particles in a homogeneous, time-independent electric field $\mathbf{E}$. Relevant observables are the occupation numbers of momentum eigenstates $\{|\mathbf{p}\rangle\}$. When applied to a relevant number operator $N(\mathbf{p})$, the evolution superoperator simply causes a shift in momentum,

$$\mathcal{U}(\tau,0) N(\mathbf{p}) = N(\mathbf{p} + q\mathbf{E}\tau) \quad , \tag{100}$$

with $q$ denoting the electric charge of the particles. Relevant observables are thus being transformed into relevant observables; relevant and irrelevant degrees of freedom are not being mixed. Therefore, independently of the choice of projector, the memory term vanishes because $\mathcal{QL}N(\mathbf{p}) = 0$. There remains only the first term

$$\begin{aligned}
\dot{n}(\mathbf{p},t) &= i(\rho(t)|\mathcal{P}(t)\mathcal{L}N(\mathbf{p})) \\
&= -\frac{\mathrm{d}}{\mathrm{d}\tau}\bigg|_{\tau=0} (\rho(t)|\mathcal{P}(t)\mathcal{U}(\tau,0)N(\mathbf{p})) \\
&= -\frac{\mathrm{d}}{\mathrm{d}\tau}\bigg|_{\tau=0} n(\mathbf{p} + q\mathbf{E}\tau, t) \\
&= -q\mathbf{E}\cdot\nabla_\mathbf{p} n(\mathbf{p},t) \quad .
\end{aligned} \tag{101}$$

This is the well-known acceleration term in the quantum Boltzmann equation. It represents an important example in which the first term of the general equation of motion does not vanish even though all relevant observables commute (cf. our remark in section 3.4.2.). Since there is no integration over past history involved, the acceleration term is always Markovian and a time scale analysis not required.

## 4.3 Rate Equation for Elastic Scattering

### 4.3.1 Non-Markovian Equation of Motion

**The underlying microscopic processes are transitions between energy levels due to a perturbing external potential.** Relevant observables are the occupation numbers of the unperturbed levels; the appropriate basis states $\{|i\rangle\}$ are thus given by the eigenstates $\{|\epsilon_0^i\rangle\}$ of the unperturbed Hamiltonian. Since the perturbing potential is assumed to be external, and not the result of particle interactions, the full Hamiltonian has the form

$$H = H^{(0)} + V = \sum_i \epsilon_0^i N_i + \sum_{kl} V^{kl} a_k^\dagger a_l \quad . \tag{102}$$

(Our notation does not account for possible level degeneracies. However, a generalization of the formalism which includes such degeneracies is straightforward.) As the total number of particles is a constant of the motion, $[H, N] = 0$, any grand canonical state

$$\rho_{\text{eq}}[\alpha, \beta] := Z^{-1} \exp(-\beta H + \alpha N) \tag{103}$$

is stationary.

**In this particular application Kawasaki-Gunton and Mori projector coincide.** Observables which appear in this model are all of the single-particle form $\sum O^{ij} a_i^\dagger a_j$. It is therefore the effect on $a_i^\dagger a_j$ which is the only interesting property of a projector $\mathcal{P}$. In this respect, the Kawasaki-Gunton projector $\mathcal{P}_{\text{KG}}[n_i(t)]$ and the Mori projector $\mathcal{P}_{\text{M}}[\alpha, 0]$, defined in $\rho_{\text{eq}}[\alpha, 0]$, are equivalent: both yield

$$\mathcal{P} a_i^\dagger a_j = \delta_{ij} N_i \quad \forall i, j \quad , \tag{104}$$

*independent* of $\{n_i(t)\}$ or $\alpha$. We may thus use a generic symbol $\mathcal{P}$ for the projector, and invoke its interpretation as either Kawasaki-Gunton or Mori projector only when it is needed or convenient.

**Evaluating the projected equation of motion yields a non-Markovian rate equation.** Application of the interaction Liouvillian $\mathcal{V}$ to a field operator $a_i^\dagger$ yields (for both bosons and fermions)

$$\mathcal{V} a_i^\dagger = \hbar^{-1} \sum_k V^{ki} a_k^\dagger \quad , \tag{105}$$

and to a relevant observable $N_i$

$$\mathcal{V} N_i = \hbar^{-1} \sum_{kl} V^{kl} (\delta_{il} a_k^\dagger a_i - \delta_{ik} a_i^\dagger a_l) \quad . \tag{106}$$

This implies $\mathcal{P}\mathcal{L}N_i = \mathcal{P}\mathcal{V}N_i = 0$, hence the first term in the equation of motion vanishes. So does the residual force term: Since the projector may be regarded as a Kawasaki-Gunton projector and since there are, by assumption, no initial correlations, the general argument applies (cf. section 3.4.2) according to which there is no residual force. Thus there remains only the memory term. Invoking now the interpretation of $\mathcal{P}$ as a Mori projector, we can take over the memory term of the Langevin-Mori equation:

$$\dot{n}_i(t) = \sum_k \int_0^{t-t_0} d\tau \, [-\gamma_i^k(\tau)] \delta n_k(t-\tau) \quad . \tag{107}$$

Here

$$\gamma_i^k(\tau) = [\overline{n}(1 \pm \overline{n})]^{-1} \langle \mathcal{V}N_k ; \mathcal{T}(0,\tau) \mathcal{V}N_i \rangle_\alpha \quad , \tag{108}$$

$$\delta n_k(t) = n_k(t) - \overline{n} \tag{109}$$

and

$$\overline{n} := \langle N_k \rangle_\alpha \quad \forall k \quad , \tag{110}$$

all expectation values and canonical correlation functions being evaluated in the stationary state $\rho_{\text{eq}}[\alpha, 0]$. Using

$$\sum_k \gamma_i^k(\tau) = 0 \quad , \tag{111}$$

the equation of motion may also be cast into the form

$$\dot{n}_i(t) = \sum_k \int_0^{t-t_0} d\tau \, [-\gamma_i^k(\tau)] \cdot [n_k(t-\tau) - n_i(t-\tau)] \quad , \tag{112}$$

a form which already looks very much like a rate equation. It is, however, still non-Markovian.

**The memory matrix is calculated in perturbation theory.** In case the perturbing potential is weak, the memory matrix may be evaluated to lowest non-trivial (i. e., second) order:

$$\gamma_i^k(\tau) = \begin{cases} (2/\hbar^2) \operatorname{Re}\left[\sum_{l \neq i} |V^{li}|^2 \exp(i\omega_{li}\tau)\right] : k = i \\ -(2/\hbar^2) \operatorname{Re}\left[|V^{ki}|^2 \exp(i\omega_{ki}\tau)\right] : k \neq i \end{cases} . \tag{113}$$

(Here we used the abbreviation $\omega_{ki} := (\epsilon_0^k - \epsilon_0^i)/\hbar$.) It is this result from second order perturbation theory which we want to subject to a time scale analysis.

*4.3.2 Memory Time*

**The memory time is related to the inverse coupling width.** For our analysis we assume that the number of levels is large; that the system is close to its equilibrium configuration $\{n_i^{\text{eq}}\}$; and, more stringently, that only one particular occupation number $n_j(t)$ is perturbed appreciably from its equilibrium value:

$$n_i(t) = \begin{cases} n_j^{\text{eq}} + \delta n_j(t) : i = j \\ n_i^{\text{eq}} \qquad\qquad : i \neq j \end{cases} . \tag{114}$$

The rate equation then describes the relaxation of $n_j(t)$ towards equilibrium:

$$\dot{\delta n}_j(t) = -\int_0^{t-t_0} d\tau\, \gamma_j^j(\tau)\, \delta n_j(t-\tau) . \tag{115}$$

From this expression one can extract both the memory time $\tau_{\text{mem}}^j$ and the relaxation time $\tau_{\text{rel}}^j$ which are associated with the occupation number $n_j$. (Note that for each occupation number there is a different set of time scales.) Information pertaining to these scales is encoded entirely in the memory matrix element $\gamma_j^j(\tau)$. Inspection of the perturbative result shows that $\gamma_j^j(\tau)$ is essentially the Fourier transform of $|V^{lj}|^2$ with respect to the frequency difference $\omega_{lj}$. Hence if, by virtue of the perturbing potential, the level $j$ couples to an entire "band" of other levels $\{l\}$, the band having a typical frequency width $\Delta_j$, then $\gamma_j^j(\tau)$ is some distribution in $\tau$ with characteristic width $1/\Delta_j$. The latter immediately determines the memory time:

$$\tau_{\text{mem}}^j \sim 1/\Delta_j . \tag{116}$$

As an example we consider electron transport in a conductor at low temperatures (cf. section 4.3.5). The electrons move with Fermi velocity $v_F$ under the influence of randomly distributed impurities. Each impurity has a typical size $\eta$, which we assume to be large compared to the Fermi wavelength. The associated potential then couples states within a frequency band of width $\Delta \sim \Delta(p^2/2m\hbar) \sim v_F \Delta k \sim v_F/\eta$. The memory time is thus $\tau_{\text{mem}} \sim \eta/v_F$, i. e., the typical time needed for an electron to pass through the interaction range of one impurity [35].

*4.3.3 Markovian Limit and Relaxation Time*

**The relaxation time is related to the total transition rate.** Provided the memory time is sufficiently small, one may take the Markovian and quasistationary limits. The general rate equation then becomes

$$\dot{n}_i(t) = \sum_k r_{ki}[n_k(t) - n_i(t)] \quad , \tag{117}$$

where we have identified the transition rates

$$r_{ki} := \int_0^\infty d\tau \, [-\gamma_i^k(\tau)] \quad , \quad k \neq i \quad . \tag{118}$$

Inserting the perturbative result for the memory matrix yields

$$r_{ki} = (2\pi/\hbar^2)\delta(\omega_{ki})|V^{ki}|^2 \quad , \quad k \neq i \quad , \tag{119}$$

in agreement with Fermi's golden rule [68]. In the Markovian limit the relaxation towards equilibrium is exponential,

$$\dot{\delta n}_j(t) = -r_j^{\text{tot}} \delta n_j(t) \quad , \tag{120}$$

with the decay constant furnished by the total transition rate

$$r_j^{\text{tot}} := \sum_{k \neq j} r_{kj} \quad . \tag{121}$$

From this we identify the relaxation time

$$\tau_{\text{rel}}^j = 1/r_j^{\text{tot}} \quad . \tag{122}$$

**The Markovian limit is justified for narrow resonances.** With hindsight we may now conclude that the Markovian limit is justified whenever the relaxation time is much larger than the memory time; or, equivalently, whenever the total transition rate is much smaller than the coupling width ("narrow resonances") [69,70]. While the memory time does not depend on the absolute strength of the potential, but only on its shape, the relaxation time increases as the potential becomes weaker. Therefore, the quality of the Markovian approximation –like the quality of perturbation theory– generally improves as $|V^{kl}| \to 0$. It is important to keep in mind, however, that despite this common feature the Markovian limit and perturbation theory represent two independent approximations.

*4.3.4 Dynamical Correlations*

**Dynamical correlations do not depend on the state of the system.** From our previous considerations we know that in the Markovian limit

$$\gamma^k_{<i}(\omega) = \begin{cases} r^{\text{tot}}_i & : k = i \\ -r_{ki} & : k \neq i \end{cases} \quad \forall \omega \quad . \tag{123}$$

Provided the total transition rate is much larger than individual rates, the memory matrix is almost diagonal; we may then set

$$\gamma^k_{<i}(\omega) \approx \delta_{ki} \cdot 1/\tau^i_{\text{rel}} \quad \forall \omega \quad . \tag{124}$$

Inserting this approximation into Eq. (77) yields the Fourier transformed dynamical correlation matrix:

$$\Xi^k_{<i}(\omega) \approx \delta_{ki} \frac{\tau^i_{\text{rel}}}{1 - i\omega\tau^i_{\text{rel}}} \quad . \tag{125}$$

This result depends only on relaxation time and frequency, but not on the thermodynamic state (temperature, density, etc.) of the system.

*4.3.5 Application: Drude Formula for the Electrical Conductivity*

**The conductivity is defined in linear response theory.** Macroscopic (normal) conductors at low temperatures owe their non-zero resistance to the presence of impurities [35]. According to the Kubo formula of linear response theory, their conductance $G$ is determined by the current-current (*I-I*) correlation function:

$$G(\omega) = \beta \langle I; \mathcal{G}_<(\omega) I \rangle_{\text{eq}} \quad . \tag{126}$$

Here $\mathcal{G}_<$ is the Green's function superoperator (which includes the effects of elastic impurity scattering), and $\beta$ denotes the inverse temperature. Instead of the global conductance one may also consider the local *conductivity tensor* $\sigma_{ik}$, which is essentially given by the autocorrelation function of the current *density* [60]. The current density, in turn, is related to the particle momenta such that

$$\sigma_{ik}(\omega) = \frac{e^2 \beta}{m^2 V} \langle \delta P_i ; \mathcal{G}_<(\omega) \delta P_k \rangle_{\text{eq}} \quad , \quad i,k = 1 \ldots 3 \quad . \tag{127}$$

Here $i,k$ label the components of the conductivity tensor; $e$ denotes the electron charge, $m$ the electron mass, $V$ the volume; and $\delta \mathbf{P}$ is defined as

$$\delta \mathbf{P} := \mathbf{P} - \langle \mathbf{P} \rangle_{\text{eq}} \quad , \tag{128}$$

with $\mathbf{P}$ denoting the operator of total momentum.

**The conductivity is related to dynamical correlations of occupation numbers.** At this point it seems very suggestive to identify the $\{P_i\}$ as "relevant observables," and to evaluate their dynamical correlation function directly with the help of Langevin-Mori theory (Eq. (77)). However, the level of description spanned by these few observables is generally not Markovian; and as a result, the evaluation of the memory matrix may pose difficult problems. In order to make use of the powerful Markovian approximation, one should rather *extend* the level of description. Provided the impurity density is sufficiently small and hence the perturbing potential sufficiently weak, we know from our previous considerations that elastic scattering of electrons off these impurities may be described by a Markovian rate equation. In this rate equation the place of the unperturbed levels $|\epsilon_0^i\rangle$ is taken by the momentum eigenstates $|\mathbf{p}\rangle$, and that of the occupation numbers $\{N_i\}$ by $\{N(\mathbf{p})\}$. It is therefore the number operators $\{N(\mathbf{p})\}$ which span the appropriate level of description: this level is Markovian, and since

$$P_i = \sum_{\mathbf{p}} p_i N(\mathbf{p}) \quad , \tag{129}$$

it contains the original level of description as a subspace. Using (75), the conductivity tensor can then be related to the dynamical correlation of occupation numbers:

$$\sigma_{ik}(\omega) = \frac{e^2 \beta}{m^2 V} \sum_{\mathbf{p}} \sum_{\mathbf{q}} \sum_{\mathbf{q}'} q_i p_k C_{\mathbf{q}\mathbf{q}'} \Xi_{<\mathbf{p}}^{\mathbf{q}'}(\omega) \quad . \tag{130}$$

**Explicit evaluation yields the Drude formula.** Since both the ordinary correlation matrix $C$ and the dynamical correlation matrix $\Xi$ are diagonal, two of the three sums immediately collapse. Furthermore, at $T \to 0$, the diagonal elements of $C$ are given by partial derivatives of the occupation numbers with respect to the Fermi energy:

$$\beta C_{\mathbf{pp}} \to \left(\frac{\partial n(\mathbf{p})}{\partial \epsilon_F}\right)_{T \to 0} \quad . \tag{131}$$

These derivatives are non-zero only on the Fermi surface. Hence –provided the impurity scattering is isotropic– we may replace

$$\Xi^{\mathbf{p}}_{<\mathbf{p}}(\omega) \to \frac{\tau}{1 - i\omega\tau} \quad , \tag{132}$$

where $\tau$ no longer depends on the momentum $\mathbf{p}$ and denotes the elastic scattering time of electrons on the Fermi surface. The remaining summation over $\mathbf{p}$ is then performed easily: Using

$$\frac{1}{V} \sum_{\mathbf{p}} p_i p_k \left(\frac{\partial n(\mathbf{p})}{\partial \epsilon_F}\right)_{T \to 0} = \delta_{ik} m \cdot n \quad , \tag{133}$$

where $n$ denotes the number density of electrons, we obtain the well-known Drude result [71–73]

$$\sigma_{ik}(\omega) = \delta_{ik} \frac{e^2 n \tau}{m} \cdot \frac{1}{1 - i\omega\tau} \quad . \tag{134}$$

## 4.4 Source Term

### 4.4.1 Non-Markovian Equation of Motion

**Our investigation is based on a simple model from quantum electrodynamics.** Strong external fields not only cause the acceleration of charged particles, but in addition may lead to the spontaneous creation of particle-antiparticle pairs. As an example we consider the spontaneous creation of $e^+e^-$-pairs in a homogeneous, time-independent electric field $\mathbf{E}$, a process often referred to as the Schwinger mechanism [74–78]. Relevant observables are the number operators

$$N_-(\mathbf{p}, m_z) := a^\dagger(\mathbf{p}, m_z) a(\mathbf{p}, m_z) \quad,$$
$$N_+(\mathbf{p}, m_z) := b^\dagger(\mathbf{p}, m_z) b(\mathbf{p}, m_z) \quad, \tag{135}$$

whose expectation values

$$n_\mp(\mathbf{p}, m_z, t) := \langle N_\mp(\mathbf{p}, m_z) \rangle(t) \tag{136}$$

describe the momentum and spin distribution of electrons and positrons, respectively. Here $a$ and $b$ denote electron and positron field operators, $\mathbf{p}$ the momentum and $m_z$ the spin component. The transport equation will have the structure

$$\dot{n}_\mp(\mathbf{p}, m_z, t) \pm q\mathbf{E} \cdot \nabla_\mathbf{p} n_\mp(\mathbf{p}, m_z, t) = \dot{n}^{\text{sou}}_\mp(\mathbf{p}, m_z, t) \quad, \tag{137}$$

with the known acceleration term (cf. section 4.2) and a new source term which accounts for spontaneous pair creation. In the following we will show how this source term can be derived from first principles.

Our starting point is the Dirac equation

$$i\hbar \frac{\mathrm{d}}{\mathrm{d}t} |\psi(t)\rangle = H |\psi(t)\rangle \tag{138}$$

with the Hamiltonian

$$H = \mathbf{p} \cdot \alpha + m\beta + qA_0 \tag{139}$$

and electrostatic potential

$$A_0(\mathbf{r}) = -\mathbf{E} \cdot \mathbf{r} \tag{140}$$

(where $q = -|e|$ for electrons). For later use we introduce some definitions: the time-dependent momentum

$$\mathbf{p}(t) := \mathbf{p} + q\mathbf{E}t \quad , \tag{141}$$

the transverse energy

$$\epsilon_\perp := \sqrt{m^2 + \mathbf{p}_\perp^2} \quad , \tag{142}$$

the total kinetic energy

$$\epsilon[\mathbf{p}(t)] := \sqrt{\epsilon_\perp^2 + p_\parallel(t)^2} \quad , \tag{143}$$

and the dynamical phase

$$\phi_{fi} := -\frac{1}{\hbar} \int_{t_i}^{t_f} dt'\, \epsilon[\mathbf{p}(t')] \quad . \tag{144}$$

In our notation "longitudinal" ($\parallel$) and "transverse" ($\perp$) always refer to the direction of the electric field.

The basis states

$$|i,\pm\rangle \equiv |\mathbf{p}(t_i), m_z, \pm\rangle \quad , \tag{145}$$

which correspond to momentum $\mathbf{p}(t_i)$, spin component $m_z$ and positive or negative energy $\pm\epsilon[\mathbf{p}(t_i)]$, evolve according to [79–82]

$$U(t_f, t_i) \begin{pmatrix} |i,+\rangle \\ |i,-\rangle \end{pmatrix} = \begin{pmatrix} \alpha_{fi} & \beta_{fi} \\ -\beta_{fi}^* & \alpha_{fi}^* \end{pmatrix} \begin{pmatrix} e^{+i\phi_{fi}} & 0 \\ 0 & e^{-i\phi_{fi}} \end{pmatrix} \begin{pmatrix} |f,+\rangle \\ |f,-\rangle \end{pmatrix} \quad . \tag{146}$$

The evolution thus mixes positive and negative energy states, with respective amplitudes $\alpha_{fi}$ and $\beta_{fi}$. These amplitudes are determined by the differential equation [83]

$$\begin{pmatrix} \dot{\alpha}_{fi} \\ \dot{\beta}_{fi} \end{pmatrix} = \frac{qE}{2} \cdot \frac{\epsilon_\perp}{\epsilon[\mathbf{p}(t_f)]^2} \begin{pmatrix} 0 & -e^{-i2\phi_{fi}} \\ e^{+i2\phi_{fi}} & 0 \end{pmatrix} \begin{pmatrix} \alpha_{fi} \\ \beta_{fi} \end{pmatrix} \tag{147}$$

with initial conditions $\alpha_{ii} = 1$ and $\beta_{ii} = 0$, and the overdot indicating differentiation with respect to $t_f$.

**The microscopic dynamics is described by a time-dependent Bogoliubov transformation.** In view of applying the projection method, the above results have to be translated into the language of field operators. Employing the shorthand notation

$$a(i, \pm) \equiv a(\mathbf{p}(t_i), m_z, \pm) \tag{148}$$

for particle field operators, and making use of the general rule

$$\mathcal{U}(t_f, t_i) a^\dagger(\psi) = a^\dagger(U(t_f, t_i)\psi) \quad , \tag{149}$$

one finds

$$\mathcal{U}(t_f, t_i) \begin{pmatrix} a^\dagger(i,+) \\ a^\dagger(i,-) \end{pmatrix} = \begin{pmatrix} \alpha_{fi} & \beta_{fi} \\ -\beta_{fi}^* & \alpha_{fi}^* \end{pmatrix} \begin{pmatrix} e^{+i\phi_{fi}} & 0 \\ 0 & e^{-i\phi_{fi}} \end{pmatrix} \begin{pmatrix} a^\dagger(f,+) \\ a^\dagger(f,-) \end{pmatrix} ; \tag{150}$$

the evolution law for $a$ follows by Hermitian conjugation. Identifying negative energy particle operators with positive energy antiparticle operators makes evident that pair creation is described by a time-dependent Bogoliubov transformation [67,84,85]. The modulus squared of the associated amplitude, $|\beta_{fi}|^2$, equals the probability for having created an $e^+e^-$-pair with final momenta $\pm\mathbf{p}(t_f)$ during the time interval $[t_i, t_f]$.

From the evolution superoperator one obtains the Liouvillian

$$\mathcal{L} = i \left.\frac{\partial}{\partial t_f}\right|_{t_f = t_i} \mathcal{U}(t_f, t_i) \tag{151}$$

which may be written as the sum

$$\mathcal{L} = \mathcal{L}_{\text{acc}} + \mathcal{V} \tag{152}$$

of a diagonal part $\mathcal{L}_{\text{acc}}$, responsible for acceleration, and an off-diagonal part $\mathcal{V}$ which is responsible for the mixture of particle and antiparticle states and hence for pair creation. With the definition

$$\dot{\beta}_{ii} := \left.\dot{\beta}_{fi}\right|_{t_f=t_i} \quad , \tag{153}$$

the latter is given by

$$\mathcal{V} \begin{pmatrix} a^\dagger(i,+) \\ a^\dagger(i,-) \end{pmatrix} = \begin{pmatrix} 0 & i\dot{\beta}_{ii} \\ -i\dot{\beta}_{ii}^* & 0 \end{pmatrix} \begin{pmatrix} a^\dagger(i,+) \\ a^\dagger(i,-) \end{pmatrix} \quad . \tag{154}$$

**The source term represents a particular example of a non-Markovian rate equation.** Comparing Eq. (154) with Eq. (105), we note that with the identification

$$i\dot{\beta}_{ii} \leftrightarrow \hbar^{-1} V^{(i,-)(i,+)} \tag{155}$$

there is a one-to-one correspondence between the microscopic dynamics of spontaneous pair creation and level transitions in an external potential. From this we infer that the source term must have the structure of a –generally non-Markovian– rate equation. Provided the electric field is sufficiently weak, $E \ll m^2/\hbar q$, then $|\dot{\beta}| \ll |\dot{\phi}|$ and hence $\mathcal{V}$ may be regarded as a small perturbation. In this case we can immediately take over both Eq. (112) and the perturbative result (113); with the sole modification, however, that the evolution superoperator

$$\mathcal{T}(0,\tau) \to \exp(i\mathcal{Q}\mathcal{L}_{\text{acc}}\mathcal{Q}\tau) \tag{156}$$

contained in the memory term now generates not only a phase factor $[\exp(i\omega_{ki}\tau)]$, but also a shift in momentum. Therefore, the correct rate equation reads

$$\dot{n}^{\text{sou}}(\mathbf{p}, m_z, \pm; t) = 2\,\text{Re} \int_0^{t-t_0} d\tau\, \dot{\beta}^*(-\tau, -\tau) e^{-i2\phi(-\tau, 0)} \dot{\beta}(0,0)$$
$$\times [n(\mathbf{p} - q\mathbf{E}\tau, m_z, \mp; t-\tau) - n(\mathbf{p} - q\mathbf{E}\tau, m_z, \pm; t-\tau)] \tag{157}$$

with $\dot{\beta}(t_f, t_i) \equiv \dot{\beta}_{fi}$ and $\phi(t_f, t_i) \equiv \phi_{fi}$. Upon identifying electron and positron occupation numbers

$$n_-(\mathbf{p}, m_z, t) = n(\mathbf{p}, m_z, +; t) \quad,$$
$$n_+(\mathbf{p}, m_z, t) = 1 - n(-\mathbf{p}, -m_z, -; t) \tag{158}$$

we obtain the source term [45]

$$\dot{n}^{\text{sou}}_{\mp}(\pm\mathbf{p}, \pm m_z, t) = 2\,\text{Re} \int_0^{t-t_0} d\tau\, \dot{\beta}^*(-\tau, -\tau) e^{-i2\phi(-\tau, 0)} \dot{\beta}(0,0)$$
$$\times [1 - n_-(\mathbf{p} - q\mathbf{E}\tau, m_z, t-\tau)$$
$$- n_+(-\mathbf{p} + q\mathbf{E}\tau, -m_z, t-\tau)] \quad. \tag{159}$$

This source term involves an integration over the entire history of the system, thus accounting for finite memory effects and rendering the evolution of the occupation numbers generally *non*-Markovian. It exhibits two characteristic time scales: (i) the memory time $\tau_{\mathrm{mem}}(\mathbf{p})$, which corresponds to the temporal extent of each individual creation process and which indicates how far back into the past one has to reach in order to predict future occupation numbers; and (ii) the production interval $\tau_{\mathrm{prod}}(\mathbf{p})$, the inverse of the production rate, which corresponds to the average time that elapses *between* creation processes and thus constitutes the typical time scale on which the occupation numbers evolve. Both time scales refer to the specific momentum $\mathbf{p}$ with which the particles are produced. Only if $\tau_{\mathrm{mem}} \ll \tau_{\mathrm{prod}}$ can memory effects be neglected and the evolution be considered approximately Markovian.

**The source term is consistent with the Schwinger formula.** For simplicity, we will assume that (i) the field is weak ($\alpha_{fi} \approx 1$), (ii) the system is dilute ($n_{\mp} \approx 0$), and (iii) $t_0 \to -\infty$. In the weak field limit the differential equation (147) implies

$$\dot{\beta}_{ff} \exp(i2\phi_{fi}) = \dot{\beta}_{fi} \tag{160}$$

which, together with the decomposition

$$\phi(-\tau, 0) = \phi(-\tau, -\infty) - \phi(0, -\infty) \quad, \tag{161}$$

reduces the source term to

$$\begin{aligned}
\dot{n}^{\mathrm{sou}}_{-}(\mathbf{p}, m_z) &= 2\,\mathrm{Re} \int_0^{\infty} \mathrm{d}\tau\, \dot{\beta}^*(-\tau, -\infty) \dot{\beta}(0, -\infty) \\
&= 2\,\mathrm{Re}\left[\beta^*(0, -\infty) \dot{\beta}(0, -\infty)\right] \\
&= \left.\frac{\mathrm{d}}{\mathrm{d}\tau}\right|_{\tau=0} |\beta(\tau, -\infty)|^2 \quad.
\end{aligned} \tag{162}$$

The amplitude $\beta(t_f, t_i)$ depends on the momentum $\mathbf{p}$ through the initial condition $\mathbf{p}(t=0) = \mathbf{p}$. In order to obtain the *total* electron production rate per unit volume, we have to consider

$$\begin{aligned}
w &:= (2/h^3) \int \mathrm{d}^3 p\, \dot{n}^{\mathrm{sou}}_{-}(\mathbf{p}, m_z) \\
&= \frac{1}{4\pi^2 \hbar^3} \int_{m^2}^{\infty} \mathrm{d}(\epsilon_{\perp}^2) \int_{-\infty}^{\infty} \mathrm{d}p_{\parallel}\, \dot{n}^{\mathrm{sou}}_{-}(\mathbf{p}, m_z) \quad.
\end{aligned} \tag{163}$$

The numerical factor 2 corresponds to the two possible spin orientations. Noticing that in the source term

$$\left.\frac{\mathrm{d}}{\mathrm{d}\tau}\right|_{\tau=0} |\beta(\tau,-\infty)|^2 = qE \left.\frac{\mathrm{d}}{\mathrm{d}p_\|(\tau)}\right|_{p_\|(\tau)=p_\|} \left|\beta[p_\|(\tau),p_\|(-\infty)]\right|^2$$
$$= qE \frac{\mathrm{d}}{\mathrm{d}p_\|} \left|\beta[p_\|,p_\|(-\infty)]\right|^2 \quad, \tag{164}$$

the integration over the longitudinal momentum becomes straightforward:

$$\int_{-\infty}^{\infty} \mathrm{d}p_\| \, \dot{n}_-^{\mathrm{sou}}(\mathbf{p},m_z) = qE \cdot |\beta(\infty,-\infty)|^2 \quad. \tag{165}$$

The modulus squared, $|\beta(\infty,-\infty)|^2$, represents the probability that for a given transverse momentum $\mathbf{p}_\perp$ a non-adiabatic transition between the energy levels $\pm\epsilon[\mathbf{p}(t)]$ will *ever* take place. It is given by the Landau-Zener formula [86–89]

$$|\beta(\infty,-\infty)|^2 = \exp\left(-\frac{\pi\epsilon_\perp^2}{\hbar qE}\right) \quad. \tag{166}$$

The subsequent integration over $(\epsilon_\perp^2)$ is then easily performed, yielding the final result

$$w = \frac{(qE)^2}{4\pi^3\hbar^2} \exp\left(-\frac{\pi m^2}{\hbar qE}\right) \quad. \tag{167}$$

This does indeed agree with the leading term in the Schwinger formula [76].

*4.4.2 Memory Time*

**The memory time combines classical and quantum mechanical time scales.**  In the weak field limit the memory time can be easily extracted from the source term (159). The factor

$$\dot{\beta}^*(-\tau,-\tau) \propto \frac{(\epsilon_\perp/qE)}{(\tau - p_\|/qE)^2 + (\epsilon_\perp/qE)^2} \tag{168}$$

constitutes a Lorentz distribution in $\tau$, centered around $p_\|/qE$ with width $\epsilon_\perp/qE$. Significant contributions to the source term thus come from times $\tau$ which are smaller than $(p_\| + \epsilon_\perp)/qE$. The typical magnitude of $p_\|$ may be inferred if one views pair creation as a tunneling process from the negative to the positive energy continuum [90] (cf. Fig. 2): the barrier between these continua has a spatial width of the order $\epsilon_\perp/qE$, inducing a corresponding

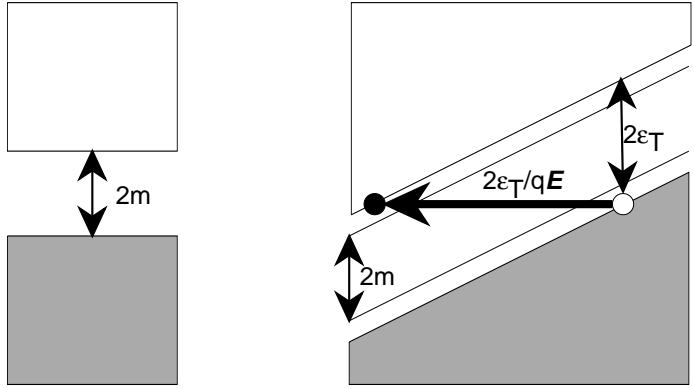

Fig. 2. The spectrum of the free Dirac-Hamiltonian is comprised of a positive and a negative energy continuum, separated by a gap of magnitude $2m$. (This gap widens to $2\epsilon_\perp$ if one fixes the transverse momentum.) In the ground state the negative energy continuum ("Dirac sea") is completely filled. Applying a homogeneous electric field "skews" these energy levels, due to the linearly rising potential. It is then possible for an electron to tunnel from the lower to the upper continuum, a process which corresponds to pair creation.

momentum scale $\Delta p_\parallel \sim \hbar q E/\epsilon_\perp$. With this typical scale one obtains

$$\tau_{\text{mem}} \sim \frac{\hbar}{\epsilon_\perp} + \frac{\epsilon_\perp}{qE} \quad . \tag{169}$$

The memory time combines two time scales of different origin: (i) The time $\hbar/\epsilon_\perp$ is proportional to $\hbar$ and therefore of quantum mechanical origin. It corresponds –via the time-energy uncertainty relation– to the time needed to create a *virtual* particle-antiparticle pair, and may thus be regarded as the "time between two production attempts." (ii) The time $\epsilon_\perp/qE$, on the other hand, is independent of $\hbar$ and therefore classical. It can be interpreted in various ways, depending on the picture employed to visualize the pair creation process. If pair creation is viewed as a tunneling process (Fig. 2), the classical memory time coincides with the time needed for the wave function to traverse the barrier with the speed of light [91]. Alternatively, pair creation may be viewed as a non-adiabatic transition between the two time-dependent energy levels $\pm\epsilon[\mathbf{p}(t)]$ (Fig. 3). In that case the classical memory time corresponds to the width of the transition region, i.e., the region of closest approach of the two levels.

For weak fields, $E \ll m^2/\hbar q$, the classical memory time dominates:

$$\frac{\epsilon_\perp}{qE} \gg \frac{\hbar}{\epsilon_\perp} \quad . \tag{170}$$

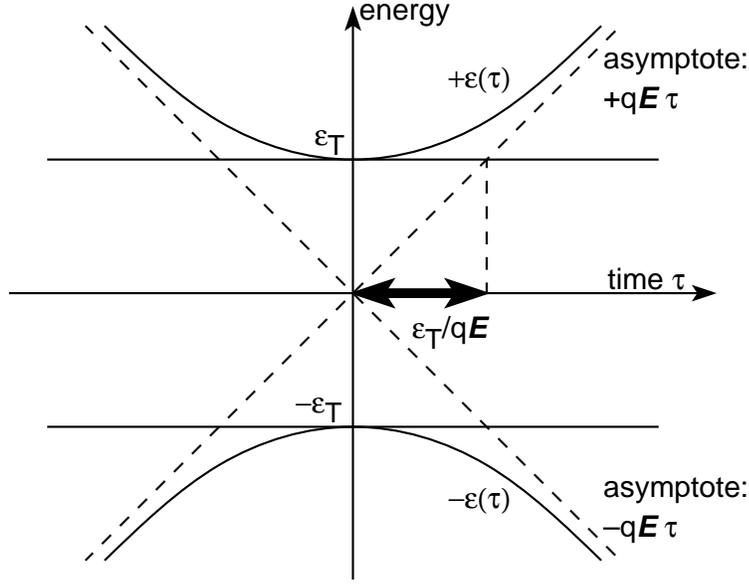

Fig. 3. Pair creation viewed as a non-adiabatic transition between two time-dependent energy levels. The temporal extent of the transition region is of the order $\epsilon_\perp/qE$.

4.4.3 *Markovian Limit and Production Interval*

**The production interval displays a characteristic $E^{-1}\exp(c/E)$-dependence.** In the weak field limit we can give a simple estimate for the order of magnitude of the production rate. According to Eq. (162)

$$\dot{n}^{\text{sou}} \sim O(|\beta\dot{\beta}|) \quad, \tag{171}$$

which together with the differential equation (147),

$$|\dot{\beta}| \sim O(qE/\epsilon_\perp) \quad, \tag{172}$$

and the Landau-Zener formula (166),

$$|\beta| \sim O\left(\exp\left(-\frac{\pi\epsilon_\perp^2}{2\hbar qE}\right)\right) \quad, \tag{173}$$

yields the estimate

$$\dot{n}^{\text{sou}} \sim O\left(\frac{qE}{\epsilon_\perp}\exp\left(-\frac{\pi\epsilon_\perp^2}{2\hbar qE}\right)\right) \quad. \tag{174}$$

Inversion gives the typical production interval

$$\tau_{\text{prod}} \sim \frac{\epsilon_\perp}{qE} \exp\left(\frac{\pi \epsilon_\perp^2}{2\hbar qE}\right) \quad . \tag{175}$$

**Pair creation in strong fields is always non-Markovian.** The evolution of occupation numbers is Markovian whenever the production interval far exceeds the memory time, $\tau_{\text{prod}} \gg \tau_{\text{mem}}$. This is the case only as long as the external field is much weaker than the critical value $m^2/\hbar q$. As soon as the field strength reaches or exceeds that critical value, the situation changes: The source term (159), which was derived in the weak-field limit, is then merely a rough estimate. Already this weak-field estimate becomes non-Markovian, indicating that at this point any conventional –i. e., Markovian– transport theory must break down. (The criteria for the validity of second order perturbation theory and of the Markovian limit are thus not independent: both approximations break down at the same critical field strength. This coincidence might have been expected, given that the model features only a single adjustable parameter $(E)$.) In other words, our analysis has revealed that in the strong-field domain pair creation must always be described by a *non-Markovian* transport theory. This non-Markovian nature may have important ramifications. For example, it may lead to oscillations of the relevant entropy, and thus temporary violations of the $H$-theorem, on the scale of the memory time (cf. appendix and Refs. [45,92]).

*4.5 Collision Term*

*4.5.1 Non-Markovian Equation of Motion*

**The underlying microscopic processes are binary collisions.** We consider the evolution of a system of identical particles (bosons or fermions) which interact through binary $(2 \to 2)$ collisions. Relevant observables are the occupation numbers of momentum eigenstates $\{|\mathbf{p}\rangle\}$. For brevity, we will often write

$$|i\rangle \equiv |\mathbf{p}_i\rangle \quad . \tag{176}$$

We assume the system to be confined to a large, but finite, volume $\Omega$ so that we can use the box normalization

$$\langle i|j\rangle = \delta_{ij} \tag{177}$$

and completeness property

$$\sum_i |i\rangle\langle i| = 1 \quad . \tag{178}$$

Our notation does not account for spin or other discrete quantum numbers. However, a generalization of the formalism which includes such additional quantum numbers is straightforward.

The microscopic Hamiltonian has the form

$$H = H^{(0)} + V = \sum_i \epsilon^i_{\text{kin}} N_i + \tfrac{1}{4} \sum_{ijkl} \langle lk|V|ji\rangle_\pm a_l^\dagger a_k^\dagger a_j a_i \quad . \tag{179}$$

Here $\epsilon^i_{\text{kin}}$ denotes the kinetic energy associated with momentum $\mathbf{p}_i$, and

$$\langle lk|V|ji\rangle_\pm := \langle lk|V|ji\rangle \pm \langle kl|V|ji\rangle \tag{180}$$

denotes an (anti-)symmetrized matrix element, the sign depending on whether the particles are bosons (upper sign) or fermions (lower sign). Defining

$$V_{lk|ji} := \langle lk|V|ji\rangle_\pm a_l^\dagger a_k^\dagger a_j a_i \tag{181}$$

with the symmetries

$$V_{lk|ji} = V_{kl|ji} = V_{lk|ij} \quad , \quad V_{lk|ji}^\dagger = V_{ij|kl} \quad , \tag{182}$$

one can also write

$$V = \tfrac{1}{4} \sum_{ijkl} V_{lk|ji} \quad . \tag{183}$$

We assume that the dynamics is invariant under spatial displacements, which implies momentum conservation:

$$V_{lk|ji} = \delta_{\mathbf{p}_i+\mathbf{p}_j,\mathbf{p}_k+\mathbf{p}_l} V_{lk|ji} \quad . \tag{184}$$

For future reference we note that

$$[V_{lk|ji}, N(\mathbf{p})] = (\delta_{\mathbf{p}_i\mathbf{p}} + \delta_{\mathbf{p}_j\mathbf{p}} - \delta_{\mathbf{p}_k\mathbf{p}} - \delta_{\mathbf{p}_l\mathbf{p}}) V_{lk|ji} \quad . \tag{185}$$

This commutator vanishes whenever $V_{lk|ji}$ corresponds to zero momentum transfer. For example, if $\mathbf{p}_l = \mathbf{p}_j$ and hence $\mathbf{p}_k = \mathbf{p}_i$ due to momentum conservation, then the Kronecker symbols add up to zero. More generally,

$$[V_{lk|ji}, N(\mathbf{p})] = 0 \quad \text{if} \quad l = j,\ l = i,\ k = j,\ \text{or}\ k = i \quad . \tag{186}$$

**We employ the Kawasaki-Gunton projector.** In this application the Kawasaki-Gunton and Mori projectors do not coincide. The choice of the Kawasaki-Gunton projector is motivated by its three potential advantages discussed in section 3.4.2: vanishing residual force, vanishing first term, and –perhaps most importantly– applicability of Wick's theorem. Thus our starting point will be the Robertson equation (63). That the first term vanishes, should be checked explicitly: Application of the interaction Liouvillian to a relevant observable $N(\mathbf{p})$ yields

$$\mathcal{V}N(\mathbf{p}) = (4\hbar)^{-1} \sum_{ijkl} [V_{lk|ji}, N(\mathbf{p})] \quad . \tag{187}$$

According to Wick's theorem, only those terms in the sum can have a non-zero expectation value $\langle\ldots\rangle_{\mathrm{rel}}$ for which $\{l, k\} = \{j, i\}$; yet in exactly that case, according to Eq. (186), the commutator vanishes. Hence no contribution survives, the first term must indeed be zero:

$$i\langle\mathcal{L}N(\mathbf{p})\rangle_{\mathrm{rel}(t)} = i\langle\mathcal{V}N(\mathbf{p})\rangle_{\mathrm{rel}(t)} = 0 \quad . \tag{188}$$

This in turn implies

$$\mathcal{P}(t)\mathcal{L}\mathcal{P}(t) = \mathcal{P}(t)\mathcal{V}\mathcal{P}(t) = 0 \quad . \tag{189}$$

In the Robertson equation, therefore, there remains only the memory term; and in this memory term all complementary projectors ($\mathcal{Q}$) may be omitted.

**The Robertson equation gives rise to a non-Markovian collision term.** We will evaluate the Robertson equation in second order perturbation theory. Combining the Robertson equation (63) with the general perturbative result (51) yields, to second order,

$$\dot{n}(\mathbf{p}, t) = -\int_0^{t-t_0} d\tau\, \langle \mathcal{V}\mathcal{U}^{(0)}(0,\tau)\mathcal{V}N(\mathbf{p})\rangle_{\mathrm{rel}(t-\tau)} \quad . \tag{190}$$

Using now Eqs. (187) and (185), as well as the property

$$\mathcal{U}^{(0)}(0,\tau)V_{lk|ji} = \exp(i\delta\omega_{lk|ji}\tau)V_{lk|ji} \tag{191}$$

with

$$\delta\omega_{lk|ji} := \hbar^{-1}(\epsilon^l_{\text{kin}} + \epsilon^k_{\text{kin}} - \epsilon^j_{\text{kin}} - \epsilon^i_{\text{kin}}) \quad , \tag{192}$$

leads to

$$\dot{n}(\mathbf{p},t) = -\frac{1}{16\hbar^2} \int_0^{t-t_0} d\tau \sum_{abcd} \sum_{ijkl} \exp(i\delta\omega_{lk|ji}\tau)(\delta_{\mathbf{p}_i\mathbf{p}} + \delta_{\mathbf{p}_j\mathbf{p}} - \delta_{\mathbf{p}_k\mathbf{p}} - \delta_{\mathbf{p}_l\mathbf{p}})$$
$$\times \langle[V_{dc|ba},V_{lk|ji}]\rangle_{\text{rel}(t-\tau)} \quad . \tag{193}$$

The expectation value in the integrand can be calculated with the help of Wick's theorem (a tedious, but straightforward calculation). One finds

$$\sum_{abcd}\langle[V_{dc|ba},V_{lk|ji}]\rangle = 4|\langle lk|V|ji\rangle_\pm|^2[n_i n_j(1\pm n_k)(1\pm n_l) - (ij \leftrightarrow kl)] \tag{194}$$

where $n_i := \langle a_i^\dagger a_i \rangle$, all expectation values being evaluated in the state $\rho_{\text{rel}}(t-\tau)$. Of the remaining four summations (over $i,j,k,l$) one more can be performed, yielding finally the non-Markovian collision term

$$\dot{n}(\mathbf{p},t) = \hbar^{-2}\text{Re}\int_0^{t-t_0} d\tau \sum_{\mathbf{p}_1\mathbf{p}_2\mathbf{p}_3} f[\mathbf{p}\mathbf{p}_3|\mathbf{p}_2\mathbf{p}_1](\tau)\cdot C[\mathbf{p}\mathbf{p}_3|\mathbf{p}_2\mathbf{p}_1](t-\tau) \tag{195}$$

with collision bracket

$$C[\mathbf{p}\mathbf{p}_3|\mathbf{p}_2\mathbf{p}_1](t') := n(\mathbf{p}_1,t')n(\mathbf{p}_2,t')[1\pm n(\mathbf{p}_3,t')][1\pm n(\mathbf{p},t')]$$
$$-(\mathbf{p}_1\mathbf{p}_2 \leftrightarrow \mathbf{p}_3\mathbf{p}) \tag{196}$$

and prefactor

$$f[\mathbf{p}\mathbf{p}_3|\mathbf{p}_2\mathbf{p}_1](\tau) := \exp(i\delta\omega[\mathbf{p}\mathbf{p}_3|\mathbf{p}_2\mathbf{p}_1]\tau)\cdot|\langle\mathbf{p}\mathbf{p}_3|V|\mathbf{p}_2\mathbf{p}_1\rangle_\pm|^2\delta_{\mathbf{p}+\mathbf{p}_3,\mathbf{p}_2+\mathbf{p}_1} \quad . \tag{197}$$

Here we have inserted an extra Kronecker symbol to make the conservation of momentum explicit. The frequency increment $\delta\omega$ which appears in the phase factor is essentially the net "energy gain" in a collision $1+2 \to 3+X$. Of course, it must vanish if the collision conserves energy. Energy conservation, however, has not yet been established. Therefore the phase factor $\exp(i\delta\omega\ldots)$ may not be neglected; in fact, it will play a crucial role in the analysis of time scales.

*4.5.2 Memory Time*

**For our analysis we make some simplifying assumptions.** (i) The dynamics is nonrelativistic; hence for particles with mass $m$,

$$\delta\omega[\mathbf{p}\mathbf{p}_3|\mathbf{p}_2\mathbf{p}_1] = \frac{1}{2m\hbar}(\mathbf{p}^2 + \mathbf{p}_3^2 - \mathbf{p}_2^2 - \mathbf{p}_1^2) \quad . \tag{198}$$

Defining the momentum transfer

$$\hbar\mathbf{q} := \mathbf{p} - \mathbf{p}_1 = \mathbf{p}_2 - \mathbf{p}_3 \tag{199}$$

and the velocity change

$$\mathbf{v} := (\mathbf{p} - \mathbf{p}_2)/m = (\mathbf{p}_1 - \mathbf{p}_3)/m \quad , \tag{200}$$

one may also write

$$\delta\omega = \mathbf{q}\cdot\mathbf{v} \quad . \tag{201}$$

(ii) We assume that the interaction has a typical range $\eta$ and make the generic choice of a Gaussian potential [38]

$$V(r) \propto \exp(-r^2/\eta^2) \quad . \tag{202}$$

Neglecting the effects of (anti-)symmetrization of the matrix elements, this leads to

$$|\langle\mathbf{p}\mathbf{p}_3|V|\mathbf{p}_2\mathbf{p}_1\rangle_\pm|^2 \propto \exp(-q^2\eta^2/2) \quad . \tag{203}$$

(iii) The system is dilute so that one may neglect the enhancement/blocking factors $[1\pm n]$. (iv) The system is close to its equilibrium configuration $\{n_\text{eq}(\mathbf{p})\}$, i.e.,

$$n(\mathbf{p},t) = n_\text{eq}(\mathbf{p}) + \delta n(\mathbf{p},t) \tag{204}$$

where the deviations $\delta n$ are small. More stringently, we assume that only one particular occupation number $n(\mathbf{p}_0, t)$ is perturbed from its equilibrium value:

$$\delta n(\mathbf{p}, t) \begin{cases} \neq 0 \text{ if } \mathbf{p} = \mathbf{p}_0 \\ = 0 \text{ otherwise} \end{cases} . \qquad (205)$$

In the collision term only the deviation of the collision bracket from equilibrium, $\delta C := C - C_{\text{eq}}$, contributes. For $\mathbf{p} = \mathbf{p}_0$,

$$\delta C[\mathbf{p}_0 \mathbf{p}_3 | \mathbf{p}_2 \mathbf{p}_1](t') = -n_{\text{eq}}(\mathbf{p}_3)\, \delta n(\mathbf{p}_0, t') \quad . \qquad (206)$$

(Variations of $n(\mathbf{p}_1)$ and $n(\mathbf{p}_2)$ do not contribute because if, for example, $\mathbf{p}_1 = \mathbf{p}_0$ then $\mathbf{p}_2 = \mathbf{p}_3$ due to momentum conservation and $C = C[\mathbf{p}_0 \mathbf{p}_3 | \mathbf{p}_3 \mathbf{p}_0] = 0$.) The collision term then describes the relaxation of $n(\mathbf{p}_0, t)$ towards equilibrium:

$$\dot{\delta n}(\mathbf{p}_0, t) = -\hbar^{-2} \text{Re} \int_0^{t-t_0} d\tau \sum_{\mathbf{p}_1 \mathbf{p}_2 \mathbf{p}_3} \exp(i\delta\omega\tau) |\langle \mathbf{p}_0 \mathbf{p}_3 | V | \mathbf{p}_2 \mathbf{p}_1 \rangle_\pm|^2$$
$$\times \delta_{\mathbf{p}_0 + \mathbf{p}_3, \mathbf{p}_1 + \mathbf{p}_2}\, n_{\text{eq}}(\mathbf{p}_3)\, \delta n(\mathbf{p}_0, t - \tau) \quad . \qquad (207)$$

(v) Finally we make an assumption about the shape of the equilibrium distribution. Again, we neglect the effect of (anti-)symmetrization and choose a Maxwell-Boltzmann distribution with inverse temperature $\beta$:

$$n_{\text{eq}}(\mathbf{p}_3) \propto \exp(-\beta p_3^2 / 2m) \quad . \qquad (208)$$

**The memory time corresponds to the typical duration of one individual collision.** From the relaxation equation (207) one can extract both the memory time $\tau_{\text{mem}}(\mathbf{p}_0)$ and the relaxation time $\tau_{\text{rel}}(\mathbf{p}_0)$, associated with the occupation number $n(\mathbf{p}_0, t)$. Strictly speaking, there is a different set of time scales $\{\tau_{\text{mem}}(\mathbf{p}), \tau_{\text{rel}}(\mathbf{p})\}$ for each observable $n(\mathbf{p}, t)$. We will be content, however, with performing the analysis in the special case $\mathbf{p}_0 = 0$. Other memory and relaxation times can be found with the same procedure. In fact, it can be shown –but will not be shown here– that any $\tau_{\text{mem}}(\mathbf{p})$ can never be greater than $\tau_{\text{mem}}(0)$ [38]. Hence $\tau_{\text{mem}}(0)$ is actually an upper limit for the various memory times, and it is fair to regard $\tau_{\text{mem}}(0)$ as *the* memory time of the system: $\tau_{\text{mem}}(0) \equiv \tau_{\text{mem}}$. Likewise, the relaxation time $\tau_{\text{rel}}(0)$ may be regarded as representative. And although by definition $\tau_{\text{mem}}$ and $\tau_{\text{rel}}$ only apply to small deviations from equilibrium, they are nevertheless useful measures for the typical time scales on which the relevant and irrelevant degrees of freedom evolve, even in situations further removed from equilibrium.

In order to evaluate the relaxation equation explicitly, we (i) assume that the volume $\Omega$ is large, permitting to take the continuum limit

$$\sum_{\mathbf{p}_i} \to \frac{\Omega}{h^3} \int d^3 p_i \quad , \quad \delta_{\mathbf{p}_i \mathbf{p}_j} \to \frac{h^3}{\Omega} \delta^3(\mathbf{p}_i - \mathbf{p}_j) \quad ; \tag{209}$$

(ii) absorb the $\delta^3$-function into the integration over $\mathbf{p}_3$; (iii) substitute $(\mathbf{p}_1, \mathbf{p}_2) \to (\mathbf{q}, \mathbf{v})$; and (iv) use $\mathbf{p} = \mathbf{p}_0 = 0$. Up to a normalization factor we obtain

$$\dot{\delta n}(0,t) \propto \mathrm{Re} \int_0^{t-t_0} d\tau \int d^3 v \int d^3 q \, \exp\left[ i\mathbf{q} \cdot \mathbf{v}\tau - \frac{1}{2}q^2\eta^2 - \right.$$
$$\left. - \beta \left( \frac{\hbar^2 q^2}{2m} + \frac{mv^2}{2} + \hbar \mathbf{q} \cdot \mathbf{v} \right) \right] \delta n(0, t-\tau) \quad . \tag{210}$$

The integration over $\mathbf{v}$ amounts to Fourier transforming a Gaussian of width $(\beta m)^{-1/2}$, which yields another Gaussian distribution in $(\tau + i\hbar\beta)q$ of width $(\beta m)^{1/2}$. One is left with

$$\dot{\delta n}(0,t) \propto \mathrm{Re} \int_0^{t-t_0} d\tau \int d^3 q \, \exp\left[ -\frac{q^2}{2} \left( \frac{(\tau + i\hbar\beta)^2}{\beta m} + \eta^2 + \frac{\hbar^2 \beta}{m} \right) \right] \delta n(0, t-\tau)$$
$$\propto \mathrm{Re} \int_0^{t-t_0} d\tau \, (\tau^2 + 2i\hbar\beta\tau + \eta^2 \beta m)^{-3/2} \, \delta n(0, t-\tau) \quad . \tag{211}$$

Any normalization factors have been ignored, since we are only interested in the form of the integrand's $\tau$-dependence. Clearly, the integrand exhibits two characteristic time scales, $(\eta\sqrt{\beta m})$ and $(\hbar\beta)$. Times $\tau$ which are greater than both of these characteristic scales yield only a negligible contribution to the integral. In other words, only those times $\tau$ contribute significantly which are smaller than either $(\eta\sqrt{\beta m})$ or $(\hbar\beta)$. This finding may be summarized in a compact fashion by setting

$$\tau_{\mathrm{mem}} \sim \eta\sqrt{\beta m} + \hbar\beta \quad . \tag{212}$$

The two contributions to the memory time have very different origins:

- The time $\tau_{\mathrm{mem}}^{\mathrm{class}} := \eta\sqrt{\beta m}$ is independent of $\hbar$ and therefore of entirely classical origin. Taking into account that $\beta \propto (m\langle v^2 \rangle)^{-1}$, one finds that

$$\tau_{\mathrm{mem}}^{\mathrm{class}} \propto \eta / \sqrt{\langle v^2 \rangle} \tag{213}$$

is just the average time needed for a particle to pass through an interaction range. During that time correlations between colliding particles may not be neglected. The classical memory time depends both on the form of the interaction (through $\eta$) and on the particle distribution (through $\beta$).

– The time $\tau_{\text{mem}}^{\text{qu}} := \hbar\beta$, on the other hand, is proportional to $\hbar$ and therefore of purely quantum mechanical origin. It is obviously a remnant of the coherent superposition of individual $2 \to 2$ scattering processes. More precisely, it describes the average temporal extent of a quantum mechanical scattering process associated –via the time-energy uncertainty relation– with the characteristic energy scale $\beta^{-1}$ of the system. For shorter times $\tau < \hbar\beta$ energy conservation is not yet established, the colliding particles are still off mass shell, and hence any subsequent collision would not be independent of the first; in short, for $\tau < \hbar\beta$ a scattering process cannot be regarded as completed. This quantum mechanical memory time ("off-shell time") is independent of the interaction and depends only on the particle distribution (through $\beta$).

In summary, the memory time corresponds to the typical duration –both classical and quantum mechanical– of a single binary collision. As in the case of level transitions, the memory time does not depend on the strength, but only on the *form* (range $\eta$) of the interaction. There is an additional feature, however, in that the memory time now also depends (through $\beta$) on the particle distribution. Both the classical and the quantum mechanical memory time decrease with increasing temperature. However, their respective temperature-dependence differs: $\sqrt{\beta}$ vs. $\beta$. There is a "critical" temperature $T_c$ at which both memory times are of equal importance:

$$kT_c \sim \hbar^2/m\eta^2 \quad . \tag{214}$$

For lower temperatures, $T < T_c$, the quantum mechanical memory time dominates; while for higher temperatures, $T > T_c$, the classical memory time prevails. Example: For a dilute gas of interacting nucleons ($mc^2 \sim 1$ GeV, $\eta \sim 1$ fm) this "critical" temperature is $kT_c \sim 40$ MeV.

We finally mention that the above result for the memory time agrees with an analysis by Danielewicz who, instead of the projection method, employed Green's function techniques [38]. Earlier discussions of the collision term can be found in, e. g., Refs. [27,93].

*4.5.3 Markovian Limit and Relaxation Time*

**The relaxation time corresponds to the average time that elapses between two successive collisions.** Provided the memory time is sufficiently small, one can take the Markovian limit $\delta n(\mathbf{p}_0, t - \tau) \to \delta n(\mathbf{p}_0, t)$, as well as the quasistationary limit $(t - t_0) \to \infty$. The $\tau$-integration in the relaxation equation (207) is then easily performed, yielding

$$\mathrm{Re} \int_0^\infty \mathrm{d}\tau \, \exp(i\delta\omega\tau) = \pi\delta(\delta\omega) \quad . \tag{215}$$

The above approximations thus imply energy conservation in each individual collision. According to Fermi's golden rule (119) the transition rate for scattering $(\mathbf{p}_3, \mathbf{p}) \to (\mathbf{p}_1, \mathbf{p}_2)$ is given by

$$r[\mathbf{p}\mathbf{p}_3|\mathbf{p}_2\mathbf{p}_1] = (2\pi/\hbar^2)\delta(\delta\omega) \, |\langle \mathbf{p}\mathbf{p}_3|V|\mathbf{p}_2\mathbf{p}_1 \rangle_\pm|^2 \delta_{\mathbf{p}+\mathbf{p}_3, \mathbf{p}_1+\mathbf{p}_2} \quad . \tag{216}$$

Taking the sum over all final states $(\mathbf{p}_1, \mathbf{p}_2)$, dividing by the incident flux $|\mathbf{v}_3 - \mathbf{v}|/\Omega$, and multiplying by an extra factor $1/2$ to account for the identity of the colliding particles, yields the total cross section

$$\sigma_{\mathrm{tot}}[\mathbf{p}, \mathbf{p}_3] = \frac{\Omega}{|\mathbf{v}_3 - \mathbf{v}|} \cdot \frac{1}{2} \sum_{\mathbf{p}_1 \mathbf{p}_2} r[\mathbf{p}\mathbf{p}_3|\mathbf{p}_2\mathbf{p}_1] \quad . \tag{217}$$

With this expression for the total cross section the relaxation equation takes on the simple form

$$\dot{\delta n}(\mathbf{p}_0, t) = -\frac{1}{\Omega} \left( \sum_{\mathbf{p}_3} |\mathbf{v}_3 - \mathbf{v}_0| \, \sigma_{\mathrm{tot}}[\mathbf{p}_0, \mathbf{p}_3] \, n_{\mathrm{eq}}(\mathbf{p}_3) \right) \delta n(\mathbf{p}_0, t) \quad . \tag{218}$$

It has the structure of an exponential decay law. For $\mathbf{p}_0 = \mathbf{v}_0 = 0$, the decay rate is given by

$$\begin{aligned} \Gamma &= \frac{1}{\Omega} \sum_{\mathbf{p}'} |\mathbf{v}'| \, \sigma_{\mathrm{tot}}[0, \mathbf{p}'] \, n_{\mathrm{eq}}(\mathbf{p}') \\ &\equiv n \langle v\sigma \rangle \quad , \end{aligned} \tag{219}$$

where $n := N/\Omega$ denotes the particle density and $\langle \ldots \rangle$ the equilibrium average. Associated with this decay rate is the relaxation time

$$\tau_{\mathrm{rel}} = \frac{1}{n \langle v\sigma \rangle} \quad . \tag{220}$$

The relaxation time is intimately connected with the mean free path $\lambda_f = 1/n\sigma$; in fact, it is nothing but the *average time between two successive collisions*.

**The Markovian limit is justified whenever the average time between collisions is much larger than the temporal extent of each individual scattering process.** The Markovian limit becomes exact as $\mathcal{V} \to 0$ (and hence $\sigma \to 0$) or $n \to 0$; i.e., for weakly interacting or very dilute systems. In contrast, the quality of the Markovian limit does not generally improve with increasing temperature, since the average velocity cancels from the ratio $\tau_{\text{mem}}^{\text{class}}/\tau_{\text{rel}}$. It might improve nevertheless if –as is often the case– the cross section decreases with increasing temperature.

# 5 Conclusion

In the preceding sections we have demonstrated the use of the projection method in several practical applications. Following the agenda laid out in section 4.1, we have employed the projection technique to investigate how a variety of physical processes, such as level transitions, binary collisions or spontaneous pair creation, manifest themselves in macroscopic transport equations. In all these examples we have ascertained that the projection method permits the derivation of macroscopic transport equations from the underlying microscopic dynamics; that the thus obtained transport equations are generally not Markovian, but exhibit a finite memory; that these non-Markovian equations may be subjected to a rigorous time scale analysis, yielding criteria for the validity of the Markovian limit; and that, where appropriate, the Markovian limit then allows one to recover conventional –i. e., Markovian– quantum transport equations. These findings substantiate the claims made in the introduction (section 1) about the advantages of the projection method. Furthermore, they confirm the general statement made in section 3.1: namely, that the mapping of the influence of irrelevant degrees of freedom onto a temporal non-locality, and the resultant possibility of analyzing and exploiting well-separated time scales, are the primary reasons for the success of the projection method.

In concluding, we wish to point out some current problems in quantum transport theory which we consider worth investigating and which, as we believe, may lend themselves to a treatment with projection techniques:

- the formulation of a transport theory for highly excited many-body systems that goes beyond the conventional Boltzmann description by accounting for memory effects, the existence of several particle species, and the possibility of cluster formation (i. e., non-trivial correlations);
- the study of dissipative phenomena and of possible memory effects in effective quantum field theories;
- the study of the transport properties of quantum systems whose classical counterparts are chaotic; and
- a better understanding of the phenomenon of decoherence.

This list of potential projects is certainly influenced by our own research interests and by no means complete. We hope that our review will help stimulate investigations both of the above and of many other related issues.

**Acknowledgement**

Parts of this review grew out of lectures given by one of us (J.R.) at the University of Debrecen. J.R. would like to thank Kornél Sailer for his kind hospitality, and him and his students for their constructive criticism. Financial support by the Heidelberger Akademie der Wissenschaften and by the U.S. Department of Energy (grant no. DE-FG05-90ER40592) is gratefully acknowledged.

## A  Entropy and Irreversibility

### A.1  Entropy and Relative Entropy

**The entropy characterizes the lack of information as to a system's microstate.** Let $A$ be a (classical, not quantum) random experiment with possible outcomes $\{a_i\}$ and respective probabilities $\{p(a_i)\}$. An "ignorance measure" $I[A]$, which characterizes the lack of information as to the outcome of $A$, should have the following three properties:

(i) $I[A]$ is a continuous, symmetric function of the probabilities $\{p(a_i)\}$.
(ii) The lack of information is non-negative, $I[A] \geq 0$. Furthermore, $I[A]$ vanishes only if the outcome of the experiment is certain, i. e., if $\{p(a_i)\} = \{0,\ldots,0,1,0,\ldots\}$.
(iii) The lack of information is additive. Let $B$ be a second experiment with possible outcomes $\{b_j\}$, and $A \cap B$ the combined experiment with outcomes $\{a_i \cap b_j\}$. Then the total lack of information, $I[A \cap B]$, is the sum of $I[A]$ and the expected lack of information as to $B$, given the outcome of $A$:

$$I[A \cap B] = I[A] + \sum_i p(a_i) I[B|a_i] \quad . \tag{A.1}$$

Here the measure $I[A \cap B]$ is defined in terms of $\{p(a_i \cap b_j)\}$, $I[A]$ in terms of $\{p(a_i)\}$, and $I[B|a_i]$ in terms of the conditional probabilities $\{p(b_j|a_i)\}$, the various probabilities being related via Bayes' rule

$$p(a_i \cap b_j) = p(a_i) p(b_j|a_i) \quad . \tag{A.2}$$

According to a celebrated theorem by Shannon [94–96] the above properties specify the ignorance measure uniquely (up to a prefactor):

$$I[A] = -k \sum_i p(a_i) \ln p(a_i) \quad , \quad k > 0 \quad . \tag{A.3}$$

Let $\{i\}$ denote a collection of discrete classical microstates of a physical system and $\{p_i\}$ their respective probabilities. The physical *entropy*

$$S[p] := -k \sum_i p_i \ln p_i \quad , \tag{A.4}$$

with $k$ chosen to be the Boltzmann constant, is then nothing but the amount of missing information as to the outcome of a most accurate experiment (i. e., an experiment which can resolve individual microstates).

The above definition of the entropy can be extended to quantum systems. Any statistical operator $\rho$ may be written as an incoherent mixture of pure states,

$$\rho = \sum_i p_i \, |i\rangle\langle i| \quad . \tag{A.5}$$

The lack of information as to the system's micro- (or pure) state is then given by $S[p]$, which in turn coincides with the *von Neumann entropy*

$$S[\rho] := -k \, \mathrm{tr}(\rho \ln \rho) \quad . \tag{A.6}$$

**For continuous probability distributions the entropy is not well defined.** For the entropy of a continuous (classical) probability distribution $p(x)$ one could envision the following definition which, however, turns out to be inappropriate: One might discretize the distribution by dividing the support manifold into bins $\{\Delta_i\}$, set

$$p_i := \int_{x \in \Delta_i} \mathrm{d}x \, p(x) \quad , \tag{A.7}$$

and define

$$S_{\mathrm{naive}}[p] := \lim_{\mathrm{vol}(\Delta_i) \to 0} \left( -k \sum_i p_i \, \ln p_i \right) \quad . \tag{A.8}$$

Yet the limit is ill-defined: this "naive" entropy diverges. And even after subtracting the divergent contribution $(-k \ln[\mathrm{vol}(\Delta_i)])$, the finite expression

$$S'_{\mathrm{naive}}[p] := -k \int \mathrm{d}x \, p(x) \ln p(x) \tag{A.9}$$

is still not acceptable, because it is not invariant under coordinate transformations. A physical information measure should not depend on the particular coordinate system in which the distribution is represented.

**A more suitable concept is the relative entropy.** Instead of asking "How much information as to the microstate is lacking?" one might also ask the reverse question: "How much information do we already have, relative to a state of total ignorance?" In the case of a discrete probability distribution the state of total ignorance corresponds to a uniform distribution $\{m_i = const.\}$. The information gained by replacing $\{m_i\}$ with a nonuniform distribution $\{p_i\}$ is then given simply by the difference of entropies, $(S[m] - S[p])$. This difference may also be written in the form

$$S_{\text{disc}}(m|p) := k \sum_i p_i \ln \frac{p_i}{m_i} \geq 0 \quad . \tag{A.10}$$

It is in the latter form that the information gain can be defined also for continuous distributions (with compact support):

$$S_{\text{cont}}(m|p) := k \int \mathrm{d}x\, p(x) \ln \frac{p(x)}{m(x)} \geq 0 \quad . \tag{A.11}$$

Except for pathological cases, this integral is well-defined and finite. Furthermore, it is invariant under coordinate transformations $x \to y$, provided not only $p$, but also $m$ is transformed:

$$\int \mathrm{d}x\, p(x) \ln \frac{p(x)}{m(x)} = \int \mathrm{d}y\, \tilde{p}(y) \ln \frac{\tilde{p}(y)}{\tilde{m}(y)} \tag{A.12}$$

where

$$p(x)\mathrm{d}x = \tilde{p}(y)\mathrm{d}y \quad , \quad m(x)\mathrm{d}x = \tilde{m}(y)\mathrm{d}y \quad . \tag{A.13}$$

In arbitrary coordinate systems the state $m$ of total ignorance no longer necessarily corresponds to a uniform distribution. The above definition of $S_{\text{cont}}(m|p)$ ensures that even in these coordinate systems the information gain is well-defined.

The preceding considerations are intended to motivate the introduction of the concept of *relative* (or Kullback) *entropy* [97,98]. The relative entropy $S(q|p)$ of a distribution $p$ with respect to another distribution $q$ is just defined as above –in both its discrete and continuous versions– but with $q$ and $p$ now representing *arbitrary* distributions. This generalizes the concept of information gain to cases in which the distribution $q$ no longer represents the state of total ignorance. The relative entropy is still non-negative, finite and coordinate-independent; it vanishes only if the two distributions coincide:

$$S(q|p) = 0 \quad \text{iff} \quad p = q \quad . \tag{A.14}$$

The definition of relative entropy can be readily extended to quantum systems. For two arbitrary statistical operators $\sigma$, $\rho$ one defines [99]

$$S(\sigma|\rho) := k\, \mathrm{tr}[\rho(\ln \rho - \ln \sigma)] \geq 0 \quad . \tag{A.15}$$

Note that in general the relative entropy is neither symmetric, $S(\sigma|\rho) \neq S(\rho|\sigma)$, nor does it equal the difference of ordinary entropies, $S(\sigma|\rho) \neq S[\sigma] - S[\rho]$.

**The relative entropy measures the distance between distributions.**
To test the quality of a statistical hypothesis, one usually employs the so-called $\chi^2$ test. The quantity $\chi^2$ characterizes how significantly the observed relative frequencies $\{p_i\}$ (data) deviate from the theoretical predictions $\{q_i\}$ (hypothesis); $\chi^2$ may thus be regarded as a measure for the *distance* between these distributions. It is defined as

$$\chi^2 := N \sum_{i=1}^{r} \frac{(p_i - q_i)^2}{q_i} \quad , \tag{A.16}$$

with $r$ denoting the number of bins (or classes) and $N$ the total number of data points (events). The distance measure $\chi^2$ is intimately connected with the relative entropy $S(q|p)$. Provided the two distributions $p, q$ are sufficiently close to each other, one may expand $S(q|p)$ in powers of $(p_i - q_i)$ and obtains, to lowest order [100],

$$S(q|p) \approx \frac{k}{2N}\chi^2 \quad . \tag{A.17}$$

This result strongly suggests that the relative entropy, too, may be regarded as a measure for the distance between distributions. Note that except for infinitesimal distances the relative entropy, like $\chi^2$, is generally not symmetric: $S(q|p) \neq S(p|q)$.

A.2 *Three Uses of Relative Entropy in Statistical Physics*

**Entropy and relative entropy serve as diagnostic tools.** Entropy and relative entropy are not physical observables like energy or particle number. In quantum mechanics there is nothing like a Hermitian "entropy operator." Rather, entropy and relative entropy serve primarily as mathematical tools to analyze and compare properties of different probability distributions. Below we will discuss three important uses of these tools: (i) finding unbiased distributions, (ii) testing the quality of approximations, and (iii) testing whether a level of description may be contracted.

*A.2.1 Finding Unbiased Prior Distributions*

**Minimizing the information gain yields a generalized canonical distribution.** Often a probability distribution is not known exactly. Rather, only certain constraints are given of the form

$$\left.\begin{array}{l} \sum_i p_i G_a(i) \\ \int \mathrm{d}x\, p(x) G_a(x) \\ \mathrm{tr}(\rho G_a) \end{array}\right\} = \langle G_a \rangle \quad, \tag{A.18}$$

the particular form depending on whether the distribution is classical discrete, continuous, or quantum. From this limited information one likes to infer that distribution $p$ which, while satisfying all constraints, is "maximally non-committal" with regard to missing information and hence "least biased." Mathematically, this requirement amounts to *minimizing*, under the given constraints, the *information gain* $S(m|p)$ relative to a state $m$ of total ignorance [101–104].

In the discrete and quantum cases, minimizing $S(m|p)$ is equivalent to maximizing the ordinary entropy $S[p]$. In the continuous case, however, there is the additional complication of finding, in the given coordinates $x$, the –possibly nonuniform– distribution $m(x)$ which represents total ignorance. This ignorance distribution $m(x)$ can usually be determined on the basis of symmetry considerations: If the structure of the support manifold is invariant under some characteristic group (e. g., translations, rotations, canonical transformations) then $m(x)\mathrm{d}x$ must coincide with the group-invariant measure [105,106]. In all cases minimizing the information gain yields a generalized canonical distribution:

$$p_i = Z^{-1} \exp(-\lambda^a G_a(i)) \quad , \quad Z = \sum_i \exp(-\lambda^a G_a(i))$$

$$p(x) = Z^{-1} m(x) \exp(-\lambda^a G_a(x)) \quad , \quad Z = \int \mathrm{d}x\, m(x) \exp(-\lambda^a G_a(x))$$

$$\rho = Z^{-1} \exp(-\lambda^a G_a) \quad , \quad Z = \mathrm{tr}\, \exp(-\lambda^a G_a) \quad . \tag{A.19}$$

(A summation over $a$ is implied.) One obvious application of these generalized canonical distributions is to equilibrium statistical mechanics, where the $\{G_a\}$ represent the constants of the motion.

*A.2.2 Testing the Quality of Approximations*

**The relative entropy measures the size of the error.** From now on we focus on quantum systems. Let the macrostate of a quantum system be characterized by the expectation values of the relevant observables $\{G_a\}$; let $\{g_a(t)\}$ be the exact expectation values, and $\{g'_a(t)\}$ some approximate values obtained after, e. g., taking the Markovian limit or doing perturbation theory. Associated with both sets of expectation values are generalized canonical states $\rho_{\text{can}}[g_a(t)]$ and $\rho_{\text{can}}[g'_a(t)]$, respectively. An appropriate measure for the quality of the approximation $g \to g'$ is furnished by the distance between these canonical states, i. e., by the relative entropy $S(\rho_{\text{can}}[g'_a(t)]||\rho_{\text{can}}[g_a(t)])$. Provided the deviations of the approximate from the exact expectation values,

$$\Delta g_a(t) := g'_a(t) - g_a(t) \quad , \tag{A.20}$$

and of the approximate from the exact associated Lagrange parameters,

$$\Delta \lambda^a(t) := \lambda^{a\,\prime}(t) - \lambda^a(t) \quad , \tag{A.21}$$

are sufficiently small, then the relative entropy is given by

$$\begin{aligned} S(\rho_{\text{can}}[g'_a(t)]||\rho_{\text{can}}[g_a(t)]) &= \tfrac{1}{2} C_{ab} \Delta \lambda^a \Delta \lambda^b + O((\Delta\lambda)^3) \\ &= \tfrac{1}{2} (C^{-1})^{ab} \Delta g_a \Delta g_b + O((\Delta g)^3) \quad . \end{aligned} \tag{A.22}$$

The relative entropy thus increases quadratically with the deviations $\Delta \lambda$ or $\Delta g$. The "metric tensor" $C$ is just the correlation matrix defined in Eq. (31). It may be evaluated (to lowest order in $\Delta\lambda$ or $\Delta g$) either in the exact or in the approximate canonical state.

*A.2.3 Testing the Feasibility of Level Contractions*

**Whether or not a level of description may be contracted, can be decided by comparing ordinary entropies.** Sometimes the characterization of the macrostate by the expectation values of $\{G_a\}$ is redundant: a *smaller* level of description,

$$\text{span}\{1, F_\alpha\} \subset \text{span}\{1, G_a\} \quad , \tag{A.23}$$

might already be sufficient, yielding approximately the same macrostate:

$$\rho_{\text{can}}[f_\alpha(t)] \approx \rho_{\text{can}}[g_a(t)] \quad . \tag{A.24}$$

If this is the case then the level of description may be *contracted* from the original (span$\{1, G_a\}$) to its subspace (span$\{1, F_\alpha\}$). An appropriate measure for the feasibility of such a contraction is again furnished by the relative entropy $S(\rho_{\text{can}}[f_\alpha(t)]||\rho_{\text{can}}[g_a(t)])$. One can show that when $\rho_{\text{can}}[f_\alpha(t)]$ is obtained from $\rho_{\text{can}}[g_a(t)]$ via level contraction, the relative entropy coincides with the difference of ordinary entropies:

$$S(\rho_{\text{can}}[f_\alpha(t)]||\rho_{\text{can}}[g_a(t)]) = S[\rho_{\text{can}}[f_\alpha(t)]] - S[\rho_{\text{can}}[g_a(t)]] \geq 0 \quad . \tag{A.25}$$

Whether or not the level contraction span$\{1, G_a\} \to$ span$\{1, F_\alpha\}$ is permitted, can thus be decided simply by comparing the respective ordinary entropies. Their difference should not exceed some prescribed upper bound which depends on the accuracy desired for the relevant expectation values (cf. the preceding section).

## A.3 Temporal Variation of Entropies

**Non-trivial statements refer to the relevant entropy.** The von Neumann entropy associated with the full statistical operator of a closed quantum system,

$$S[\rho(t)] := -k \operatorname{tr}(\rho(t) \ln \rho(t)) \quad , \tag{A.26}$$

is constant in time. Hence for the study of irreversible phenomena it is useless. Non-trivial statements all refer to the time-dependent *relevant entropy*,

$$S_{\text{rel}}[g_a(t)] := -k \operatorname{tr}(\rho_{\text{rel}}(t) \ln \rho_{\text{rel}}(t)) \quad , \tag{A.27}$$

which is the entropy associated with the relevant part of the statistical operator. It measures, at all times $t$, the amount of missing information as to the pure state of the system if one is given the current expectation values $\{g_a(t)\}$ of only the relevant observables. The transition $\rho(t) \to \rho_{\text{rel}}(t)$, and consequently $S \to S_{\text{rel}}(t)$, is often referred to as "coarse-graining." It always involves discarding information about irrelevant degrees of freedom, a truncation which is reflected in the general inequality

$$S_{\text{rel}}[g_a(t)] \geq S[\rho(t)] \quad \forall \ t \quad . \tag{A.28}$$

About the relevant entropy one can make two distinct, logically independent statements:

(i) *Second law.* Whenever both the initial and the final macrostate of an isolated system are characterized by a generalized canonical distribution –in particular, whenever the evolution of the system leads from one equilibrium state to another equilibrium state– then the associated *final* relevant entropy can never be smaller than the *initial* relevant entropy. This is true even if the initial and final levels of description differ. No statement is made about the behavior of the relevant entropy at intermediate times.

(ii) *H-theorem.* If a level of description is Markovian then its accompanying relevant entropy increases monotonically at all times. (Boltzmann's original *H*-theorem is one particular manifestation of this more general result.)

These two statements will be derived and discussed separately.

### A.3.1 Second Law

**The final relevant entropy cannot be smaller than the initial relevant entropy.** Let $\{G_a\}$ and $\{\tilde{G}_\alpha\}$ be two –possibly different– sets of relevant observables with respective expectation values $\{g_a(t)\}$, $\{\tilde{g}_\alpha(t)\}$. Let the initial state be canonical with respect to the first set $\{G_a\}$:

$$\rho(t_0) = \rho_{\text{can}}[g_a(t_0)] \quad . \tag{A.29}$$

By assumption, the relevant entropy and the von Neumann entropy are initially equal: $S_{\text{rel}}[g_a(t_0)] = S[\rho(t_0)]$. Then by unitarity, $S[\rho(t_0)] = S[\rho(t)]$, and by the general inequality for coarse-graining, $S[\rho(t)] \leq S_{\text{rel}}[\tilde{g}_\alpha(t)]$. Hence for any later time $t \geq t_0$, and for *any* choice of the second set $\{\tilde{G}_\alpha\}$ of relevant observables, it is always

$$S_{\text{rel}}[\tilde{g}_\alpha(t)] \geq S_{\text{rel}}[g_a(t_0)] \quad . \tag{A.30}$$

This is the second law in its most general form [17]. Whenever both the initial and the final macrostate of an isolated system are described by a generalized canonical distribution, possibly with respect to different levels of description, then the associated final relevant entropy cannot be smaller than the initial relevant entropy. In case the evolution of the system leads from one equilibrium state to another, the sets $\{G_a\}$ and $\{\tilde{G}_\alpha\}$ represent the "old" and "new" constants of the motion, respectively. While the second law ensures that the final relevant entropy is never smaller than the initial relevant entropy, it does not exclude local oscillations of the relevant entropy at intermediate times.

Even in the presence of such local oscillations, however, the relevant entropy never falls below its initial value.

**The second law follows from reproducibility.** As discussed lucidly by Jaynes [107], the second law is a consequence of the fundamental requirement that experiments on macroscopic systems be reproducible. In such experiments the microscopic details of the system are generally beyond the control of the experimenter; all he can control and observe is some set of macroscopic parameters. Hence the *preparation* of the system amounts to adjusting those few expectation values $\{g_a(t_0)\}$ which characterize the initial macrostate. The system is then allowed to evolve –usually until it reaches a new equilibrium– and at the end it is again only a limited set of expectation values $\{\tilde{g}_\alpha(t)\}$, namely those which characterize the final macrostate, that will be measured. The experiment is reproducible if its outcome (i. e., the final expectation values $\{\tilde{g}_\alpha(t)\}$) is uniquely determined by the initial preparation (i. e., by the initial expectation values $\{g_a(t_0)\}$); in other words, if the final $\{\tilde{g}_\alpha(t)\}$ can be predicted on the basis of the initial $\{g_a(t_0)\}$. But a prediction cannot possibly contain more information than the data on which it is based. Hence the final expectation values cannot carry more information (as to the microstate of the system) than do the initial expectation values; which implies that the corresponding final relevant entropy cannot be smaller than the initial relevant entropy, Q.E.D.

*A.3.2  H-theorem*

**On a Markovian level of description the relevant entropy increases monotonically.** If and only if the level of description spanned by the relevant observables $\{G_a\}$ is Markovian, the general argument for the second law can be applied iteratively: the expectation values $\{g_a(t+dt)\}$ at time $t+dt$ cannot carry more information than the expectation values $\{g_a(t)\}$ at time $t$, based on which they could be predicted. (A closely related argument for the $H$-theorem is sketched in figure A.1.) As a result, the associated relevant entropy can only increase or stay constant, but never decrease; in short:

$$\dot{S}_{\text{rel}}[g_a(t)] \geq 0 \quad \forall\, t \quad . \tag{A.31}$$

There is a common misconception that the $H$-theorem constitutes a proof of the second law. This is wrong. The $H$-theorem is in fact weaker than the second law in that it refers to only *one* set of relevant observables, not two; and that it holds only if the level of description is Markovian. On the other hand, it is stronger in that it describes the behavior of the relevant entropy at all times $t$, rather than just comparing initial and final values.

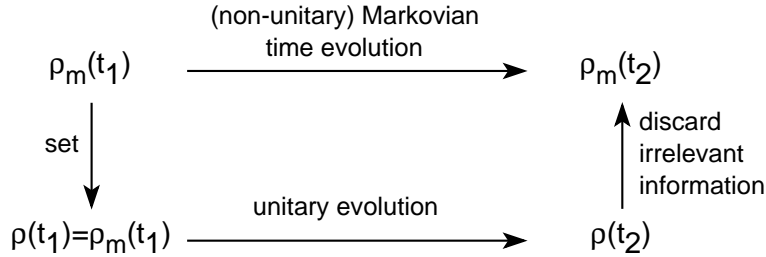

Fig. A.1. One argument for the $H$-theorem. $\rho_m$ denotes the reduced state associated with the Markovian level of description; $\rho$ denotes the full state. In the Markovian limit the diagram must be commutative. Since the lower path involves discarding information, entropy must necessarily increase: $S_m(t_2) \geq S_m(t_1)$, Q.E.D.

**Information flows from slow to fast degrees of freedom.** The $H$-theorem reveals once more the special status of the Markovian level of description: not only is the Markovian level particularly useful for explicit calculations, it is also distinguished conceptually. As we discussed before, this distinction is linked to the existence of widely disparate time scales. On the Markovian level of description "relevant" means "slow," while "irrelevant" means "fast." The growth of the Markovian relevant entropy indicates that the amount of information carried by the slowly varying observables steadily decreases. At the same time the constancy of the von Neumann entropy shows that complete information about the system is retained in a full microscopic description. An obvious interpretation is that in the course of the system's evolution, information is continuously being transferred from slow to fast degrees of freedom. It is this flow of information which is perceived as irreversible. The information is not really lost; it only becomes inaccessible to a certain coarse-grained (namely the Markovian) level of description. In the example of a dilute classical gas, information is being transferred from single-particle observables to many-particle correlations. Since in general these correlations will not be measured, part of the information about the system becomes experimentally inaccessible.

**As the Markovian entropy approaches the equilibrium entropy, the system thermalizes.** For brevity, let us denote by $\rho_m(t)$ the relevant part of the statistical operator which is associated with a Markovian level of description, and by $S_m(t)$ the associated Markovian entropy. The Markovian level of description, being spanned by all slowly varying observables, contains as a subspace the equilibrium level of description which is spanned by all constants of the motion. As a consequence, the Markovian entropy can never be greater than the equilibrium entropy:

$$S_m(t) \leq S_{\text{eq}} \quad \forall t \quad . \tag{A.32}$$

While the Markovian entropy keeps increasing due to the $H$-theorem, the equilibrium entropy is a constant. Whether or not the Markovian entropy eventually approaches the equilibrium entropy as $t \to \infty$, depends on the particular physical process. If it does, i. e., if

$$S_m(t) \to S_{\text{eq}} \tag{A.33}$$

and hence

$$S(\rho_{\text{eq}}|\rho_m(t)) \to 0 \tag{A.34}$$

then, because the relative entropy is an appropriate distance measure, one may conclude that the Markovian distribution approaches the equilibrium distribution. The level of description may then be *contracted* from the original Markovian to the equilibrium level. This means that the system's macrostate, originally described on the larger Markovian level of description, is now characterized completely by the equilibrium distribution and hence solely by the expectation values of the constants of the motion. This level contraction is the mathematical manifestation of the physical "equilibration," or "thermalization," of the system.

The above line of argument reveals the importance of the entropy concept even away from equilibrium. Being an appropriate distance measure, the entropy difference $[S_{\text{eq}} - S_m(t)]$ is the proper diagnostic tool to describe the approach to equilibrium.

**Thermalization may occur in several stages.**  Between the original ($\mathcal{M}^{(0)}$) and the equilibrium ($\mathcal{E}$) level of description there may lie several intermediate Markovian levels of description,

$$\mathcal{M}^{(0)} \supset \ldots \supset \mathcal{M}^{(i)} \supset \mathcal{M}^{(i+1)} \supset \ldots \supset \mathcal{M}^{(n)} \supset \mathcal{E} \quad, \tag{A.35}$$

each spanned by observables which evolve on successively longer time scales. Associated with these intermediate levels are reduced states $\rho_m^{(i)}(t)$ and entropies $S_m^{(i)}(t)$. The thermalization of the system then occurs in various stages, each stage amounting to a level contraction $\mathcal{M}^{(i)} \to \mathcal{M}^{(i+1)}$. Again, the entropy difference

$$[S_m^{(i+1)}(t) - S_m^{(i)}(t)] = S(\rho_m^{(i+1)}(t)|\rho_m^{(i)}(t)) \to 0 \tag{A.36}$$

is the proper diagnostic tool to test the feasibility of such a contraction. The thermalization is complete as soon as one may contract $\mathcal{M}^{(n)} \to \mathcal{E}$.

The various intermediate level contractions occur on successively longer time scales; they represent different physical "regimes" of the macroscopic evolution. Two examples are the kinetic and hydrodynamic regimes in the evolution of a dense plasma. The former is described on the Boltzmann level of description, spanned by all single-particle observables; whereas the latter is described on the hydrodynamic level of description, spanned only by the observables of local particle, energy, and momentum density. As the plasma evolves, it first reaches local equilibrium, permitting to contract the Boltzmann to the hydrodynamic level of description; then, on a generally much longer time scale, it approaches global equilibrium, eventually permitting to contract the hydrodynamic to the equilibrium level of description. These regimes and their associated time scales can be identified by studying solely the evolution of the relevant entropies $S_m^{(i)}(t)$; a fact which underlines once more the role of the entropy as a powerful diagnostic tool.

**The $H$-theorem is experimentally relevant.** One may wonder about the experimental relevance of the $H$-theorem, given that the Markovian level of description usually differs from the experimental level. However, for an observable to be measurable in practice it is usually necessary that it vary slowly. Hence the Markovian level typically *contains* the experimental level (Fig. A.2). For this case a typical behavior of the associated entropies is shown in Fig. A.3. Both the von Neumann and the equilibrium entropy remain constant in time. In contrast, the Markovian entropy increases monotonically due to the $H$-theorem. As $t \to \infty$ the Markovian entropy need not always approach the equilibrium entropy; however, if it does, then the system has thermalized. About the behavior of the experimental entropy no general statement can be made — it may or may not have local extrema. But as long as the experimental level of description is contained in the Markovian level, it is $S_m \leq S_{\exp} \leq S_{\text{eq}}$. Then thermalization on the Markovian level immediately implies thermalization on the experimental level.

That the experimental entropy may actually decrease *temporarily*, can be illustrated with a simple example [36]. Imagine, for instance, two clouds of a very dilute, very weakly interacting gas, each initially in local equilibrium, moving towards one another with large average velocities. Let us assume that only the local particle, energy, and momentum densities are measured experimentally. On this level of description the associated experimental entropy will sharply increase as soon as the two clouds begin to overlap; but after the encounter, it will *fall back* to practically its initial value. This decrease reflects the non-Markovian nature of the experimental level of description: the values of the local densities at time $t + \mathrm{d}t$ depend not only on their values at time $t$, but also on their past history before the encounter. In contrast, the Boltzmann entropy –which is associated with a Markovian level of description– will increase monotonically throughout the evolution of the system.

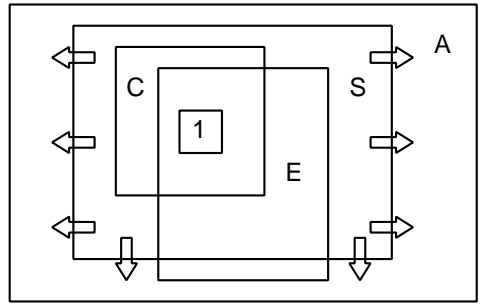

general case

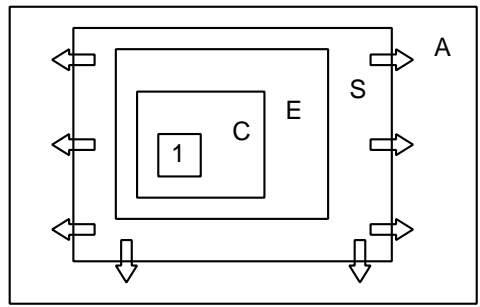

typical case

Fig. A.2. Levels of description of a physical system. 1 denotes the unit operator, $C$ the constants of the motion, $E$ the experimentally measured observables, $S$ the slow observables, and $A$ all observables. The arrows indicate the flow of information in the Markovian limit ($H$-theorem).

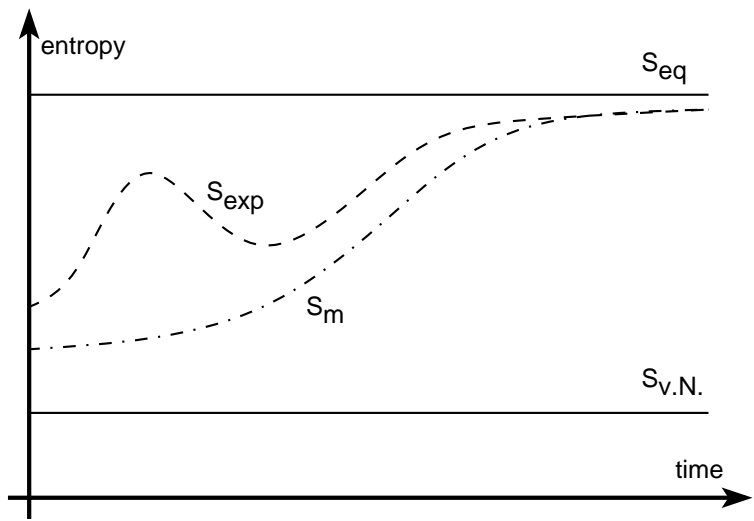

Fig. A.3. An example for the behavior of various entropies. Both the von Neumann and the equilibrium entropy remain constant in time. The Markovian entropy increases monotonically ($H$-theorem). No such general statement can be made about the behavior of the experimental entropy; it may or may not have local extrema. However, the experimental level of description is typically contained in the Markovian level, and hence $S_{\mathrm{exp}} \geq S_m$.